\documentclass[
 reprint,
 superscriptaddress,
 amsmath,
 amssymb,
 aps,
 pra,
 floatfix,
]{revtex4-2}

\usepackage{graphicx}
\usepackage{dcolumn}

\usepackage{soul}
\usepackage[normalem]{ulem}
\sethlcolor{pink}
\usepackage[caption=false]{subfig}

\usepackage[dvipsnames]{xcolor}
\usepackage{tikz}
\usetikzlibrary{arrows}
\usetikzlibrary{shapes.geometric}

\graphicspath{ {./figures/} }
\usepackage{physics}

\usepackage{url}
\usepackage{hyperref}
\hypersetup{
    colorlinks=true,
    linkcolor=blue,
    citecolor=blue, 
    filecolor=magenta,      
    urlcolor=blue,
}

\usepackage{stmaryrd}
\newcommand{\nkd}[3]{ \left \llbracket #1,\, #2,\, #3 \right \rrbracket}
\newcommand{\specialcell}[2][c]{%
  \begin{tabular}[#1]{@{}c@{}}#2\end{tabular}}
  
\definecolor{Xorange}{HTML}{F5793A}
\definecolor{Ypurple}{HTML}{A95AA1}
\definecolor{Zblue}{HTML}{85C0F9}
\definecolor{plaq}{HTML}{0F2080}

\definecolor{Mygreen}{HTML}{009E74}
\definecolor{Mydarkblue}{HTML}{0072b2}
\definecolor{Myblue}{HTML}{56b3e9}
\definecolor{Myyellow}{HTML}{f0e442}
\definecolor{Myorange}{HTML}{e69d00}
\definecolor{Myred}{HTML}{d55c00}
\definecolor{Mypink}{HTML}{cc79a7}

\definecolor{CommentGreen}{RGB}{0, 160, 0}
\definecolor{BenBlue}{RGB}{0, 0, 160}

\usepackage{mleftright} 

\newcommand{\figref}[1]{\mbox{Fig.~\ref{#1}}}
\newcommand{\tabref}[1]{\mbox{Table~\ref{#1}}}
\newcommand{\secref}[1]{\mbox{Sec.~\ref{#1}}}

\newcommand{\appref}[1]{\mbox{Appendix~\ref{#1}}}
\renewcommand{\eqref}[1]{\mbox{Eq.~(\ref{#1})}}

\newcommand{\figpanel}[2]{Fig.~\hyperref[#1]{\ref*{#1}(#2)}}
\newcommand{\figpanels}[3]{Fig.~\hyperref[#1]{\ref*{#1}(#2)-(#3)}}
\newcommand{\figpanelNoPrefix}[2]{\hyperref[#1]{\ref*{#1}(#2)}}

\newcommand{\xyz}{XYZ$^2$ }

\usepackage{siunitx}
\usepackage{nicefrac}

\begin{document}

\title{Sequential decoding of the \xyz hexagonal stabilizer code}

\author{Basudha Srivastava}
\email[]{basudha.srivastava@quantinuum.com}
\affiliation{Quantinuum, Terrington House, 13-15 Hills Rd, Cambridge, CB2 1NL, UK}
\affiliation{Department of Physics, University of Gothenburg, 41296 Gothenburg, Sweden}

\author{Yinzi Xiao}
\affiliation{Department of Computer Science, Paderborn University, 33098 Paderborn, Germany}

\author{Anton Frisk Kockum}
\affiliation{Department of Microtechnology and Nanoscience, Chalmers University of Technology, 41296 Gothenburg, Sweden}

\author{Ben Criger}
\affiliation{Quantinuum, Terrington House, 13-15 Hills Rd, Cambridge, CB2 1NL, UK}
\affiliation{Institute for Globally Distributed Open Research and Education (IGDORE)}

\author{Mats Granath}
\email[]{mats.granath@physics.gu.se}
 \affiliation{Department of Physics, University of Gothenburg, 41296 Gothenburg, Sweden}

\date{\today}

\begin{abstract}

Quantum error correction requires accurate and efficient decoding to optimally suppress errors in the encoded information. For concatenated codes, where one code is embedded within another, optimal decoding can be achieved using a message-passing algorithm that sends conditional error probabilities from the lower-level code to a higher-level decoder. In this work, we study the \xyz topological stabilizer code, defined on a honeycomb lattice, and use the fact that it can be viewed as a concatenation of a $\nkd{2}{1}{1}$ phase-flip parity check code and the surface code with $YZZY$ stabilizers, to decode the syndrome information in two steps. We use this sequential decoding scheme to correct errors on data qubits, as well as measurement errors, under various biased error models using both a maximum-likelihood decoder (MLD) and more efficient matching-based decoders. For depolarizing noise we find that the sequential matching decoder gives a threshold of \SI{18.3}{\percent}, close to optimal, as a consequence of a favorable, effectively biased, error model on the upper-level YZZY code. For phase-biased noise on data qubits, at a bias $\eta = \frac{p_z}{p_x+p_y} = 10$, we find that a belief-matching-based decoder reaches thresholds of \SI{24.1}{\percent}, compared to \SI{28.6}{\percent} for the MLD. With measurement errors the thresholds are reduced to \SI{3.4}{\percent} and \SI{4.3}{\percent}, for depolarizing and biased noise respectively, using the belief-matching decoder. This demonstrates that the \xyz code has thresholds that are competitive with other codes tailored to biased noise. The results also showcase two approaches to taking advantage of concatenated codes: 1) tailoring the upper-level code to the effective noise profile of the decoded lower-level code, and 2) making use of an upper-level decoder that can utilize the local information from the lower-level code.

\end{abstract}
\maketitle

\section{Introduction}\label{sec:introduction}

Quantum computers are expected to have a computational advantage over classical computers for certain applications~\cite{Shor1994Algorithms, Grover1997, quantumalgorithmzoo, Montanaro2016, Dalzell2023}. To achieve these advantages, quantum computers must operate with significantly lower error rates than what can currently be achieved experimentally~\cite{Fowler2012, Preskill2018, Kivlichan2020, Campbell2021}. Overcoming the adverse effects of noise requires the use of quantum error-correcting codes, that will come with an overhead in space (number of qubits) and/or time (number of gates)~\cite{Nielsen2000, Devitt2013, Terhal2015}.

One of the most prominent approaches to quantum error correction (QEC) uses topological stabilizer codes~\cite{Kitaev2003} which are constructed by tiling a manifold. The number of logical qubits in these codes is determined by the topology of the manifold, while the code distance is determined by the density of the tiling. Topological codes have several favorable properties, such as requiring only local entangling gates, having low-depth syndrome measurement circuits, and high threshold error rates (below which lower logical error rates can be achieved by increasing code distance). Among these, the surface code~\cite{Fowler2012} is the most well known and has seen a growing number of experimental demonstrations in recent years~\cite{Andersen2020, Satzinger2021, Marques2021, Krinner2022, Acharya2023Suppressing, Acharya2025}. 

Another approach to achieving fault tolerance uses concatenated codes --- errors at the physical level are suppressed using an error-correcting code, and the resulting less noisy logical qubits are then re-encoded into higher-level codes. This approach offers several advantages, including better architecture when using short- and long-range gates~\cite{Criger2016Noise}, constant space overhead in the absence of connectivity constraints~\cite{Yamasaki2024Time}, and constant-factor reductions in space overhead in planar architectures~\cite{Gidney2023Yoked}, among others~\cite{Pattison2023Hierarchical, Satoshi2024Concatenate, Vasmer2021Morphing, Goto2024Hypercube, Li2023, Meister2024}.

Due to the intrinsic noise bias found in various qubit architectures, another line of recent research is the design of QEC codes tailored to such noise models~\cite{Ataides2021XZZX, Hastings2021, Haah2022, Wootton2015, Srivastava2022xyzhexagonal, Huang2023, Wootton2022}. In this work, we focus on the \xyz code~\cite{Wootton2015, Wootton2022, Srivastava2022xyzhexagonal}, a topological stabilizer code with favorable properties under biased noise. The \xyz code has local stabilizers and a code distance which is linear in the number of physical qubits for pure $Z$ noise. It has previously been shown to have high thresholds and highly suppressed logical failure rates for phase-biased noise, under the assumption of perfect stabilizer measurements~\cite{Srivastava2022xyzhexagonal}.

\begin{figure}
\includegraphics[width=0.9\linewidth]{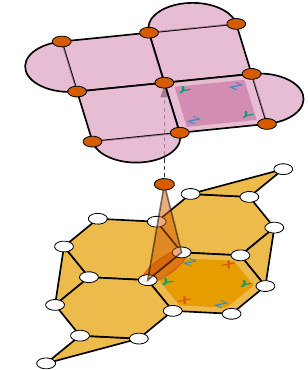}
\caption{The \xyz code can be viewed as a concatenation of the YZZY code with each data qubit in the higher-level YZZY code encoded in a lower-level $\nkd{2}{1}{1}$ phase-flip error-detecting code, with a single $XX$ stabilizer. The figure depicts the transformation from an $XYZXYZ$ plaquette stabilizer to a $YZZY$ plaquette stabilizer by replacing each pair of physical qubits (white circles) on a link (orange ellipse) with a logical qubit (orange circle) of the $\nkd{2}{1}{1}$ code.} \label{fig:xyz2_transformation}
\end{figure}

In this work, we use the fact that  \xyz code can be described as a concatenation of a $\nkd{2}{1}{1}$ phase-flip error-detecting code embedded into a surface code where each plaquette of four qubits is stabilized by the product of Pauli matrices  $YZZY$ (see \figref{fig:xyz2_transformation}), here referred to as the YZZY code~\cite{Ataides2021XZZX}. The latter is an example of a Clifford-deformed surface code~\cite{Ataides2021XZZX,Tuckett2018, Tuckett2019, Tuckett2020, Dua2024,Xiao2024Exact}, which in itself has favorable properties for phase-biased noise.  The concatenated structure allows us to implement sequential decoders for the \xyz code using Poulin's message-passing technique ~\cite{Poulin2006Optimal}. 

We make use of matching~\cite{pymatchingv2}, belief-matching~\cite{Criger2018,Higgott2023ImprovedDecoding}, and matrix-product-state-based~\cite{Bravyi2014EfficientCode,qecsim,Xiao2024Exact} decoders for the upper-level (YZZY) code under code-capacity noise (where errors may occur on data qubits, but not measurement results) and phenomenological noise (where errors occur on data qubits and measurement results independently). We find that a standard matching-based decoder that uses the conditional logical error probabilities from the lower-level syndrome measurement performs nearly optimally for code-capacity depolarizing noise. This is a consequence of a resulting biased noise profile on the upper-level YZZY code, for which the latter is well suited. For phase-biased noise on the \xyz code, the noise on the higher-level code is effectively $X$-biased. Here, pre-processing with belief propagation provides better input probabilities for the matching decoder.

In previous work~\cite{Srivastava2022xyzhexagonal}, the \xyz code was shown to have high thresholds and low logical failure rates under code-capacity noise and with near optimal decoding. Here we extend these conclusions to include measurement errors and to the use of more efficient matching-based decoders. 
The results also show how the noise profile is transformed by decoding the lower-level code of a concatenated code and the value of  having either a higher-level code which is tailored to such noise or a decoder that can utilize the information from the lower-level code.

The structure of this article is as follows. In \secref{sec:background}, we review stabilizer codes, focusing specifically on the surface code and the \xyz code. In \secref{sec:methods}, we describe the sequential decoding scheme and the noise models used to numerically simulate the performance of the \xyz code. In \secref{sec:results}, we present the results with code-capacity and phenomenological noise simulations under depolarizing and phase-biased noise. We conclude in \secref{sec:conclusion}. In \appref{app:bence-decoder}, we compare the performance of our decoder to a similar work that also decodes the \xyz code~\cite{hetenyi2024tailoring}. In \appref{app:bm-vs-matching}, we comment on the performance of a belief matching-based decoder compared to a matching decoder for a particular noise model. Finally, in \appref{app:proof-optimal-decoding}, we give the proof that the sequential decoding scheme is optimal for the \xyz code when using an optimal upper-level decoder.

\section{Background}
\label{sec:background}

Stabilizer codes are a class of quantum error-correcting codes whose logical codewords can be expressed as simultaneous $+1$ eigenstates of an Abelian subgroup $S$ of the $n$-qubit Pauli group $\mathcal{P}_n$~\cite{Gottesman1997Stabilizer, Gottesman2009Introduction}, which is generated by $\{i\} \cup \{X, Z\}^{\otimes n}$, where $X$ and $Z$ are Pauli matrices. Error correction can be performed on these codes by repeatedly measuring the stabilizers and using the set of $-1$ outcomes, known as the \textit{syndrome}, to detect and correct errors that may have occurred due to external noise. This syndrome is interpreted by a decoder, which then returns a correction that, ideally, brings the state of the system back into the code space without creating a logical error. 

A wide variety of decoding algorithms have been formulated for this purpose~\cite{deMarti_iOlius_2024}.
A multitude of decoder algorithms have been developed, ranging from minimum weight perfect matching \cite{Dennis2002,wang2009threshold, pymatchingv2}, belief-propagation (BP)-based  \cite{Criger2018, Poulin2008iterative, Roffe2020}, Markov chain Monte Carlo \cite{Wootton2012, Bravyi2014, Hutter2014,Hammar2022}, matrix-product state (MPS) contractions \cite{Bravyi2014, chubb2021generaltensornetworkdecoding}, and more~\cite{deMarti_iOlius_2024}. 
Recently, decoders based on modern deep learning algorithms have been found to outperform all other decoders for experimentally realistic noise models on the surface code~\cite{Lange2023data,Varbanov2023Neural,Bausch2023}. Here we focus on leveraging the concatenated structure of the \xyz code, with standard decoders using matching, BP-augmented matching, and MPS. 


\subsection{Surface codes}
\label{subsec:surface-code}

\begin{figure}
\includegraphics[width=0.8\linewidth]{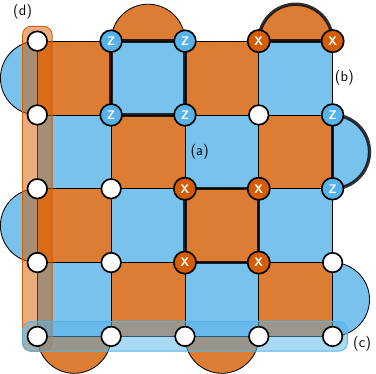}
\caption{The rotated surface code with distance $d=5$ and open boundary conditions. The physical qubits lie on the vertices of the square grid (depicted by white circles). 
(a) Orange and blue plaquettes denote the $XXXX$ and $ZZZZ$ stabilizers respectively. 
(b) Boundary stabilizers are weight-2 operators of the form $XX$ and $ZZ$. 
(c) The logical operator $Z_L = Z^{\otimes d}$ runs along a horizontal chain of qubits. 
(d) The logical operator $X_L = X^{\otimes d}$ runs across a vertical chain of qubits.}\label{fig:surface-code}
\end{figure}

The surface code is a stabilizer code with local stabilizer generators, similar to Kitaev's toric code~\cite{Dennis2002, Kitaev2003}, but with open boundaries that permit it to be used in planar devices. In this work, we will look at the rotated version of the surface code, including explicit boundaries (see \figref{fig:surface-code})~\cite{Wen2003QuantumOrders, Bombin2007}. The rotated surface code encodes a single logical qubit using a grid of $d \times d$ physical qubits, where $d$ is the distance of the code. The code's stabilizer generators are weight-4 $XXXX$ or $ZZZZ$ plaquette stabilizers in the bulk [see \figpanel{fig:surface-code}{a}], and weight-2 $XX$ or $ZZ$ stabilizers on the boundary [see \figpanel{fig:surface-code}{b}].

The minimum-weight logical operators of the code are given by chains of Pauli operators going vertically or horizontally across the code block, equivalent up to the action of the stabilizers [see \figpanels{fig:surface-code}{c}{d} for a representation of the operators $Z_L$ and $X_L$]. The likelihood of a logical error occurring depends on the rates at which different Pauli errors occur, and hence on the noise bias. When considering biased noise, we differentiate among the different logical operators based on the weights of their individual Pauli operators, and consider the distance of the code under pure (involving only one type of Pauli operator) Pauli errors~\cite{Xiao2024Exact, Higgott2023ImprovedDecoding}. For a $d \cross d$ rotated surface code, the pure $X$ distance is $d$, the pure $Z$ distance is $d$, and the pure $Y$ distance is $d^2$.

\begin{figure}
\includegraphics[width=0.8\linewidth]{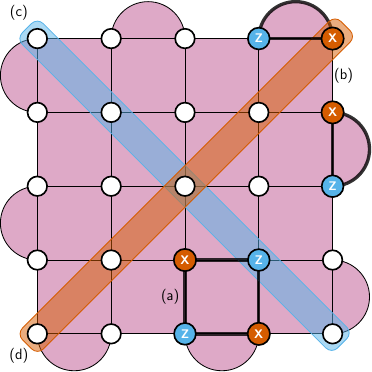}
\caption{The XZZX code with distance $d=5$ and open boundary conditions. The physical qubits lie on the vertices of the square grid (depicted by white circles). 
(a) The plaquette stabilizers are uniformly $XZZX$. 
(b) Boundary stabilizers are weight-2 $XZ$ or $ZX$. 
(c) The logical operator $Z_L = Z^{\otimes d}$ runs from the top left to the bottom right of the lattice. 
(d) The logical operator $X_L = X^{\otimes d}$ runs from bottom left to top right.}
\label{fig:xzzx-code}
\end{figure}

A simple Clifford transformation of the rotated surface code transforms it into a code where the weight-4 plaquette operators are uniformly $XZZX$~\cite{Wen2003QuantumOrders, Kay2011Capabilities, Ataides2021XZZX} (see \figref{fig:xzzx-code}). The logical operators of the code are also transformed accordingly. The distance of the code under pure errors $Z$ remains $d$, since the logical operator on the diagonal from top left to bottom right remains unchanged [see \figpanel{fig:xzzx-code}{c}]. Similarly, the diagonal from bottom left to top right supports a pure $X$ logical operator [see \figpanel{fig:xzzx-code}{d}]. Similarly to the rotated surface code, for a $d \cross d$ XZZX surface code, the pure $X$ distance is $d$, the pure $Z$ distance is $d$, and the pure $Y$ distance is $d^2$.

As noted in Ref.~\cite{Ataides2021XZZX}, pure $X$ or $Z$ Pauli noise acting on the XZZX code results in a decoding problem that can be solved using a simple algorithm similar to a repetition code decoder.
As a result, the XZZX code also exhibits high thresholds under highly biased error models~\cite{Ataides2021XZZX}. Another Clifford transformation to the rotated surface code gives the YZZY code, equivalent to the XZZX code with similar properties.

\subsection{The \xyz hexagonal code}
\label{subsec:xyz2}

\begin{figure}
\includegraphics[width=0.75\linewidth]{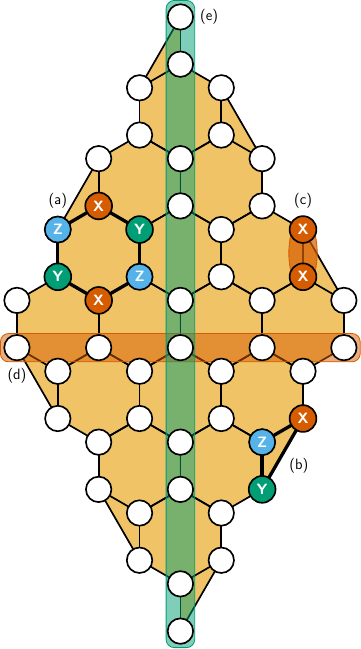}
\caption{The \xyz code with distance $d=5$. The physical qubits are depicted using white circles on the vertices of the grid. (a) The plaquette stabilizers are equal to $XYZXYZ$ on every plaquette. (b) The boundary stabilizers are equal to $XYZ$. (c) Each nearest-neighbour pair of qubits connected by a vertical edge has an additional weight-2 $XX$ link stabilizer. (d) The logical operator $X_L$ is $X^{\otimes d}$ on the $d$ qubits crossing the lattice horizontally. Note that this operator has degeneracy $2^d$, due to the $d$ $XX$ link stabilizers incident on the operator. (e) The logical operator $Y_L$ is $\{YZ\}^{\otimes 5}$ crossing the lattice vertically in the center.}\label{fig:xyz2}
\end{figure}

The \xyz code~\cite{Srivastava2022xyzhexagonal} is a stabilizer code implemented on a hexagonal grid of data qubits. With specific boundary conditions, it encodes a single logical qubit in $2d^2$ physical qubits, where $d$ is the distance of the code. The stabilizer generators are mixed Pauli weight-6 $XYZXYZ$ plaquette stabilizers, weight-3 $XYZ$ boundary stabilizers, and weight-2 $XX$ link stabilizers on nearest-neighbour pairs of qubits connected by vertical edges [see \figpanels{fig:xyz2}{a}{c}]. 

This code has a minimum-weight all-$X$ logical operator, which we will call $X_L$, supported on $d$ qubits, crossing the lattice horizontally [see \figpanel{fig:xyz2}{d}]. Similarly, there exists a minimum-weight all-$Z$ operator $Z_L$ which acts on all the data qubits. This operator is equivalent to the all-$Y$ chain, up to $XX$ link stabilizers. The third logical operator, $Y_L$, can be chosen as the vertical operator of mixed $YZ$ chains [see \figpanel{fig:xyz2}{e}]. So, for the \xyz code with $2d^2$ physical qubits, the pure $X$ distance is $d$, and the pure $Y$ or $Z$ distance is $2d^2$.

In previous work~\cite{Srivastava2022xyzhexagonal}, it was shown that the \xyz code has code-capacity thresholds close to those of the XZZX code~\cite{Ataides2021XZZX} under biased noise. In addition to these high thresholds, the \xyz code also has suppressed logical failure rates compared to the XZZX code for $Z$-biased noise close to the threshold, due to the quadratic distance ($2d^2$) for pure $Z$ noise (compared to the linear distance $d$ for the XZZX code).

\section{Methods}\label{sec:methods}

\subsection{Sequential decoding}\label{subsec:decoding}
Concatenating codes provides a means of achieving reduced error probabilities, where the logical qubits of error-correcting code blocks are encoded into another error-correcting code. Concatenated codes can be decoded optimally using a message-passing algorithm given by Poulin~\cite{Poulin2006Optimal}, which is efficient if the number of logical qubits in each layer is small. The algorithm works by decoding the lowest layer of encoded blocks and passing the conditional logical error probabilities as input to the decoder for the next higher-level code. The message-passing decoder is optimal when the errors between physical-level code blocks are independent, provided that each layer is decoded optimally. 

The \xyz code can be formulated as a concatenation of the YZZY surface code with $\nkd{2}{1}{1}$ error-detecting codes (see \figref{fig:xyz2_transformation}). We design our decoding algorithm as a two-step process based on this description, as detailed below for both code-capacity and phenomenological noise.   

In related work, Het\'enyi et al.~\cite{hetenyi2024tailoring} used a matching decoder on the \xyz code. The matching approach in that work is based on duplicating the link syndromes into two, that match separately onto the even and odd row sublattices of plaquettes. 
We present a comparison to that decoder for code-capacity noise in \appref{app:bence-decoder}.

\subsubsection{Code-capacity noise}
\label{subsec:ccn-decoding}

\begin{figure*}
\includegraphics[width=\linewidth]{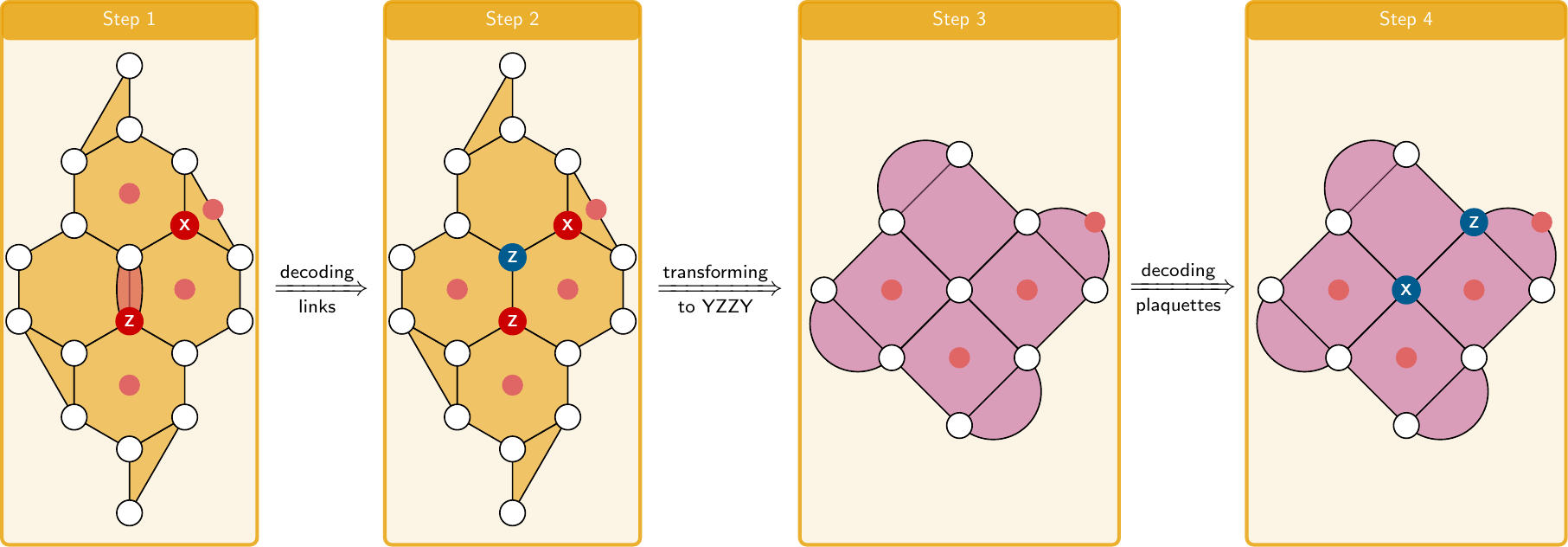}
\caption{Decoding the \xyz code as a concatenated code, assuming perfect stabilizer measurements. Step 1 takes as input the measured plaquette and link stabilizers of the \xyz code. The errors (red Pauli operators) give rise to the syndromes marked with red. Step 2 is to decode the link syndromes by acting on the top qubit of every link with a non-trivial syndrome with a Pauli $Z$ (blue Pauli $Z$) and modifying the plaquette syndromes accordingly. This step removes all non-trivial link syndromes. Step 3 transforms the code to the YZZY code. Step 4 decodes the plaquette syndromes on the YZZY code using an efficient matching/belief-matching decoder. The resulting error chain can now be transformed back to an error chain on the \xyz code, and the final correction, a product of the high-level correction and the low-level correction derived in Step 2, can be returned.
}\label{fig:decoding_2d}
\end{figure*}

The first step of our decoding algorithm, for code-capacity noise, is to decode the link syndromes of the \xyz code. An example showing the decoding steps for this is illustrated in \figref{fig:decoding_2d}. A $Z$ error on a link leads to a non-trivial link syndrome. The syndrome is degenerate with respect to the error being placed on the lower or upper qubit on the link. This is a heralded erasure from the computational subspace of the upper-level code. 




We apply a Pauli $Z$ on the `top' qubit of any triggered link syndrome. This step constrains all the link syndromes to have `+1' eigenvalue, and hence takes all the two-qubit codes back to their respective logical code spaces. The syndromes on the plaquettes are modified according to the Pauli $Z$ corrections from the previous step and the priors used to decode the higher-level code are updated according to the low-level logical error likelihoods given in \tabref{table:prior-update-rule}. With the updated priors and plaquette syndromes, the higher-level code can now be decoded using an appropriate decoder. The predicted error chain given by the decoder can be mapped back to the \xyz code and, along with the added Pauli $Z$ on the links, gives the correction. This scheme gives an optimal decoder given an optimal higher-level decoder, as outlined in \appref{app:proof-optimal-decoding}.

\begin{table}
\begin{tabular}{ c | c | c }
 \centering
 link syndrome $s$ & $s = 0$ & $s = 1$ \\
 \hline \hline
 YZZY & \multicolumn{2}{c}{\xyz} \\
 \hline
 $\Bar{I}$ & \specialcell[c]{$II \text{ or } XX$\\\textcolor{Myred}{$P_{I,0}$=$\dfrac{p_i^2+p_x^2}{p_{(s = 0)}}$}} & \specialcell[c]{$ZI \text{ or } YX$\\\textcolor{Myred}{$P_{I,1}$=$\dfrac{p_zp_i+p_xp_y}{p_{(s = 1)}}$}} \vspace{2pt} \\
 \hline
 $\Bar{X}$ & \specialcell[c]{$ZZ \text{ or } YY$\\\textcolor{Myred}{$P_{X,0}$=$\dfrac{p_z^2+p_y^2}{p_{(s=0)}}$}} & \specialcell[c]{$IZ \text{ or } XY$\\\textcolor{Myred}{$P_{X,1}$=$\dfrac{p_zp_i+p_xp_y}{p_{(s=1)}}$}} \vspace{2pt} \\
 \hline
  $\Bar{Y}$ & \specialcell[c]{$ZY \text{ or } YZ$\\\textcolor{Myred}{$P_{Y,0}$=$\dfrac{2 p_y p_z}{p_{(s=0)}}$}} & \specialcell[c]{$IY \text{ or } XZ$\\\textcolor{Myred}{$P_{Y,1}$=$\dfrac{p_yp_i+p_xp_z}{p_{(s=1)}}$}} \vspace{2pt} \\
  \hline
 $\Bar{Z}$ & \specialcell[c]{$XI \text{ or } IX$\\\textcolor{Myred}{$P_{Z,0}$=$\dfrac{2 p_xp_i}{p_{(s=0)}}$}} & \specialcell[c]{$YI \text{ or } ZX$\\\textcolor{Myred}{$P_{Z,1}$=$\dfrac{p_yp_i+p_xp_z}{p_{(s=1)}}$}}
\end{tabular}
 \caption{The set of possible error configurations on a link. The first entry refers to the top qubit of the link, where a $Z$ correction is placed in the sequential decoding scheme given a link syndrome. The four classes are denoted by the corresponding single Pauli error on the high-level YZZY model. The probabilities of the corresponding error configurations are given in orange underneath each entry.}
 \label{table:prior-update-rule}
\end{table}

In our simulations, we use the MPS decoder~\cite{qecsim} as a maximum-likelihood decoder on the YZZY code. We assign individual non-IID error probabilities to the data qubits of the YZZY code, according to whether a link syndrome has been corrected on that qubit or not. We also use the efficient, though sub-optimal, matching~\cite{pymatchingv2} and belief-matching~\cite{Criger2018} decoders on the higher-level code.  

\subsubsection{Phenomenological noise}
\label{subsec:pn-decoding}

For multiple syndrome measurements, the first step of the proposed decoding algorithm is again to decode the link syndromes of the code. However, in the case of phenomenological noise (which models syndrome measurements as error-prone, failing with a probability determined by the quantum architecture), link syndromes can occur due to measurement errors. The links are therefore first decoded in the time direction, over the set rounds of stabilizer measurements. For each of the $d^2$ links, the stabilizer (detector) measurements for $d$ rounds are decoded using a matching decoder.

\begin{figure}
\includegraphics[width=0.75\linewidth]{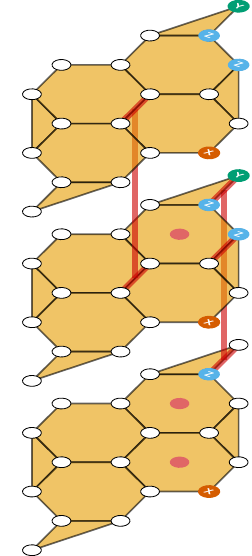}
\caption{Decoding the links of the $d=3$ \xyz code against phenomenological noise, with $d$ rounds of syndrome measurements. Each layer shows the Pauli errors that occurred at that time step, with time increasing from bottom to top. The detection events, corresponding to syndrome changes, are shown for plaquette and link stabilizers. We first solve independent decoding problems for each of the $d^2$ links in the time direction, assigning link-detection events to measurement errors or data-qubit errors based on the relative likelihood of the two cases.}\label{fig:decoding_3d}
\end{figure}

A pair of matched link syndromes can be caused by a string of measurement errors, occurring with likelihood $q^r$, where $q$ is the syndrome measurement error rate and $r$ is the distance between the two matched link detectors. Such an event pair could also be caused by two uncorrelated data errors at different time steps, with probability $p^2$, where $p$ is the physical qubit error rate. In the decoding of the links, a decision is made whether to keep the link syndromes (if $p^2 > q^r$) or remove the matched links from the next step of decoding (if $q^r > p^2$). An example showing this decoding step is illustrated in \figref{fig:decoding_3d}.

In the next step, the links of each round of stabilizer measurement are decoded using the same method as for code-capacity noise, and the plaquette syndromes are modified according to the placement of the Pauli $Z$ corrections on the qubits. The priors on the data qubits are updated accordingly. This final step of decoding the link syndromes ensures that all the link syndromes are now removed (constrained to become $+1$) and the only syndromes that remain are from the plaquette stabilizers. 

This leaves us with a 3D decoding problem for the YZZY code with non-IID underlying qubit error probabilities. In the final step, this higher-level YZZY code is decoded for the same multiple rounds of syndrome measurement using a matching or belief-matching decoder for phenomenological noise. Note that, in contrast to the algorithm described for code-capacity noise, the algorithm used for phenomenological noise is not guaranteed to be optimal. 

\subsection{Biased noise}
\label{subsec:noise-models}

In several qubit architectures, the physical noise is \textit{biased} towards a certain type of Pauli errors~\cite{Shulman2012, Pop2014, Waldherr2014, Watson2018, Lescanne2020, Hnggli2020, Hajr2024, Cong2022, Puri2020}. Using an appropriate code for such biased noise can yield higher thresholds and lower logical failure rates. Codes and decoders tailored towards exploiting the biased noise structure of the error model are thus useful for implementation on such devices~\cite{Acharya2023Suppressing, Acharya2025, Ataides2021XZZX, Tuckett2018, Tuckett2019, Tuckett2020, Dua2024, Darmawan2021, Miguel2023, Guillaud2019, Claes2023}.

The \xyz code is one such code that performs better under biased noise. It has been shown to have high thresholds for phase-flip-biased code-capacity noise.
In this work, we focus on simulating the \xyz code under a noise bias parameterized by
\begin{equation}
    \eta = \frac{p_z}{p_x+p_y}.
\end{equation}
The noise bias $\eta$ describes how much more likely a $Z$ error is to occur than an $X$ error or a $Y$ error. For example, depolarizing noise corresponds to $\eta=\nicefrac{1}{2}$, while $\eta = \infty$ signifies pure $Z$ noise. (We do not consider $\eta < \nicefrac{1}{2}$ in this work, which represents a high probability of $X$ or $Y$ errors and a low probability of $Z$ errors.)

In contrast to the earlier work~\cite{Srivastava2022xyzhexagonal}, which looked at code-capacity noise using a Monte Carlo-based maximum-likelihood decoder~\cite{Hammar2022}, here we implement matching-based decoders, with and without belief propagation, and compare to maximum-likelihood decoding using an MPS decoder~\cite{qecsim}. We also consider phenomenological noise as a na\"{i}ve but common approximation to stabilizer measurements that are not error-free in practice. For simulations, we assume a syndrome-measurement error rate of $q = p$, where $p$ is the data-qubit error rate, both for plaquette and link stabilizers.

\subsubsection{Mapping the noise bias}
\label{subsec:bias-mapping}

\begin{figure}\centering
\includegraphics[width=\linewidth, trim={0cm 0cm 1.6cm 1.4cm},clip]{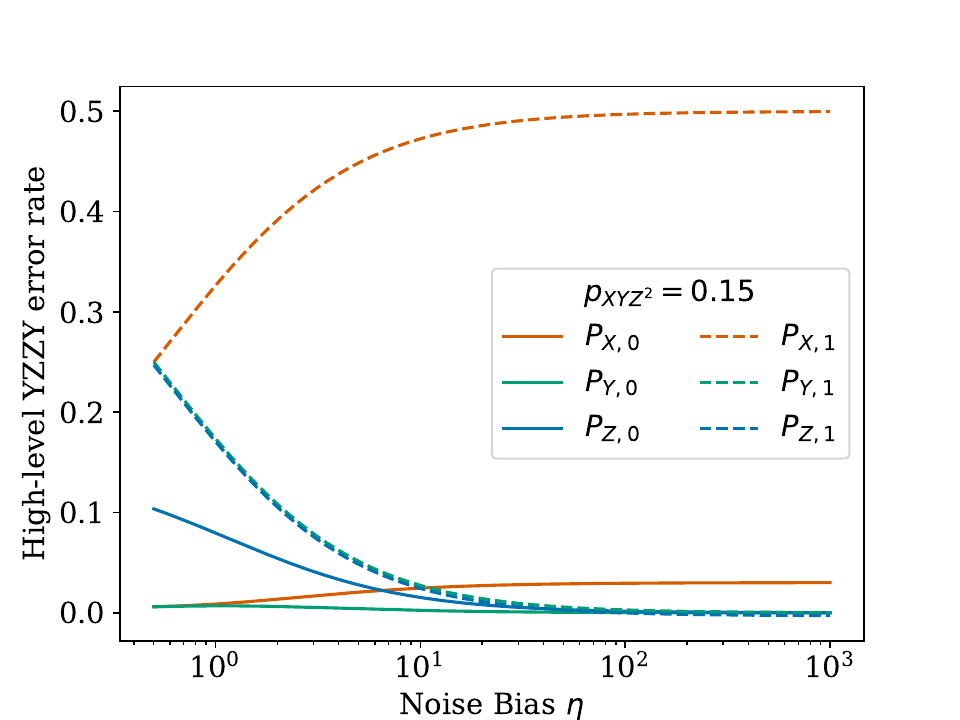}
\caption{Logical error rate of the links forming the data qubits of the YZZY code, as a function of the noise bias on the \xyz code. The plot shows $Z$-biased noise parametrized by $\eta$, at a particular physical error rate of $p_{\text{\xyz}} = 0.15$. The solid lines show the mapped error rates in the case when the link is not triggered, and the dashed lines show the mapped error rates when the links are triggered and corrected by placing a $Z$ correction on the top qubit of the link. The probabilities $P_{\sigma, 0/1}$ correspond to the expressions given in \tabref{table:prior-update-rule}.}
\label{fig:noise-bias-mapping}
\end{figure}

Given a particular error rate for the physical qubits, the error rates for the logical qubits of the repetition code on each link depend on the noise bias $\eta$. In other words, the noise model we correct with the higher-level YZZY code is modified by decoding the links according to the steps given in the previous section. This is shown in \figref{fig:noise-bias-mapping}. There, we see how the error rates used to decode the YZZY code change with the bias of the physical-level noise model, both for the YZZY-code qubits created on links with a non-triggered syndrome (given by $P_{\sigma,0}$), and for qubits made up of links with a triggered syndrome measurement (given by $P_{\sigma,1}$), where $\sigma$ is a Pauli operator.

For depolarizing noise, $\eta = \nicefrac{1}{2}$, we see that the noise bias of the non-triggered links changes to $Z$-biased noise on the YZZY code, whereas noise on the qubits from a triggered link is completely depolarizing. As $Z$-biased noise on the YZZY code gives rise to higher thresholds~\cite{Ataides2021XZZX}; we observe a similar trend. In particular, we find that this effective bias on the higher level makes a matching decoder more effective for depolarizing noise than on the non-concatenated YZZY (or surface) code. As $\eta$ increases, the noise on the YZZY code changes to increasingly $X$-biased. This can be less effectively decoded using a  matching decoder as it gives a high concentration of errors that give rise to four triggered stabilizers, i.e., a hyperedge in the syndrome graph. 

\section{Results}
\label{sec:results}

\begin{figure}\centering
\includegraphics[width=\linewidth, trim={0.5cm 0.6cm 0.58cm 0.5cm},clip]{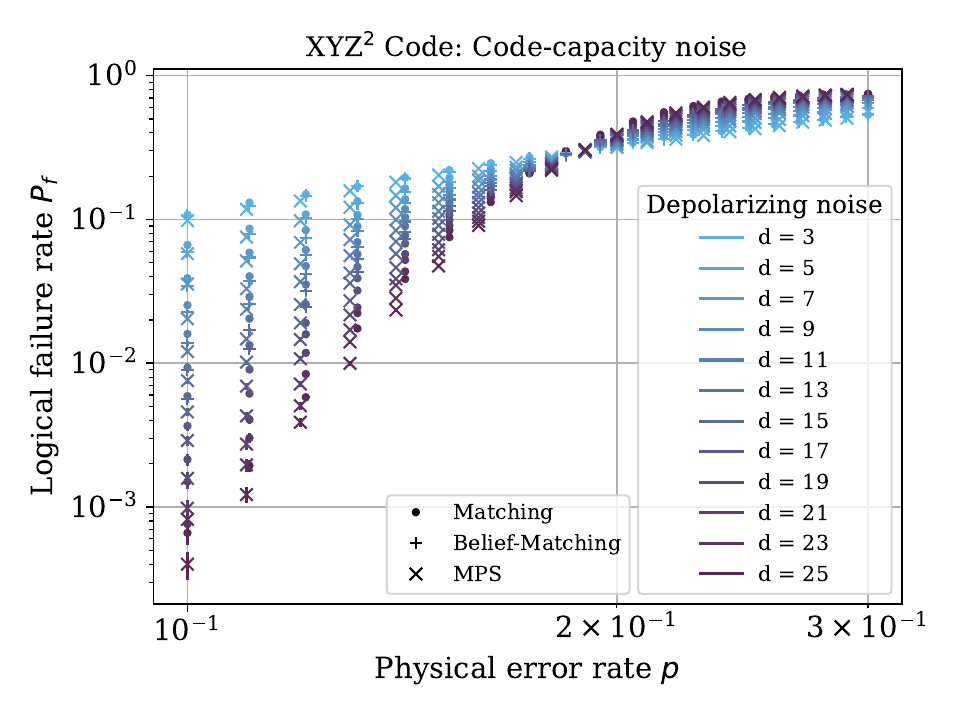}
\caption{Logical failure rate $P_f$ as a function of the physical error rate $p$ for different distances of the \xyz code, under a code-capacity depolarizing noise error model ($\eta = 0.5$). Each data point is evaluated using $5 \times 10^4$ syndromes, decoded using a matching (circular marker), belief-matching (plus marker), or MPS (cross marker) decoder on the higher-level YZZY code. For depolarizing noise on the \xyz code, the noise model transforms into $Z$-biased noise on the YZZY code, which has a high threshold using a matching decoder.  We observe thresholds of $18.34 \pm \SI{0.08}{\percent}$ and $18.52 \pm \SI{0.08}{\percent}$ using the matching and belief-matching decoder, respectively, compared to $18.92 \pm \SI{0.05}{\percent}$ for the MPS decoder. Here, and elsewhere in the text unless otherwise specified, the uncertainties on threshold error rates are calculated using the covariance matrix returned by the fitting function.}
\label{fig:code-capacity-matching-dep}
\end{figure}

We now present numerical results of our simulations of the sequential decoding strategy for the \xyz code. We first simulate the code-capacity noise model, decoding the links according to the method presented in \secref{subsec:ccn-decoding} and using the MPS, matching, and belief-matching decoders on the higher-level code. Figure~\ref{fig:code-capacity-matching-dep} shows the results for depolarizing noise on the physical qubits of the \xyz code. As explained in \secref{subsec:bias-mapping}, the noise model is transformed into $Z$-biased noise on the higher-level YZZY code. As expected from the fact that the YZZY code is known to have high thresholds for such bias using matching~\cite{Ataides2021XZZX}, we find logical error rates close to the optimal values from the MPS decoder. Correspondingly, the belief-matching decoder does not provide a significant advantage over the matching decoder.

\begin{figure}\centering
\includegraphics[width=\linewidth, trim={0.5cm 0.6cm 0.58cm 0.5cm},clip]{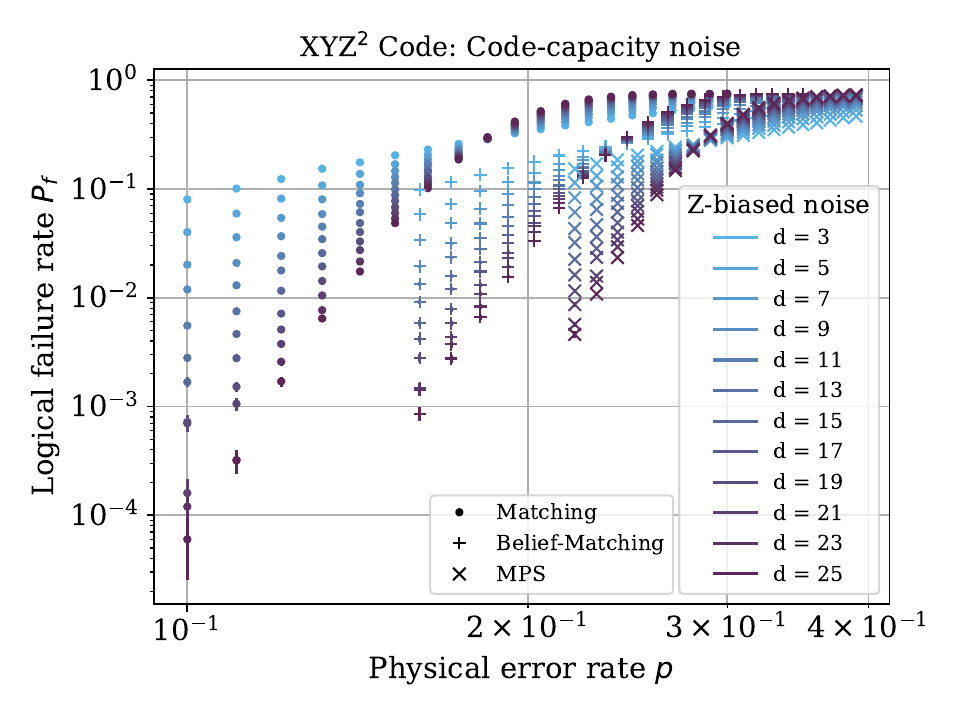}
\caption{Logical failure rate $P_f$ as a function of the physical error rate $p$ for different distances of the \xyz code, under a code-capacity $Z$-biased noise error model ($\eta = 10$). Each data point is evaluated using $5 \times 10^4$ syndromes, decoded using a matching (circular marker), belief-matching (plus marker), or MPS (cross marker) decoder on the higher-level YZZY code. We observe thresholds of $18.36 \pm \SI{0.11}{\percent}$ for the matching decoder and $24.08 \pm \SI{0.05}{\percent}$ using a belief-matching decoder, compared to $28.61 \pm \SI{0.05}{\percent}$ for the MPS decoder.}
\label{fig:code-capacity-matching-z10}
\end{figure}

In Figure~\ref{fig:code-capacity-matching-z10}, we show code-capacity logical failure data for a $Z$-biased noise channel with $\eta = 10$. The threshold difference between the matching decoder and the optimal (MPS) decoder is large. The effective noise model on the YZZY code is $X$-biased  (see \secref{subsec:bias-mapping}) which is not well represented by the two decoupled matching graphs of the matching decoder. In this instance, however, the belief-matching decoder increases the threshold and gives lower logical failure rates with $d$ rounds of belief propagation, with a threshold significantly closer to that of the MPS decoder.

\begin{figure}\centering
\includegraphics[width=\linewidth, trim={0.5cm 0.6cm 0.58cm 0.5cm}, clip]{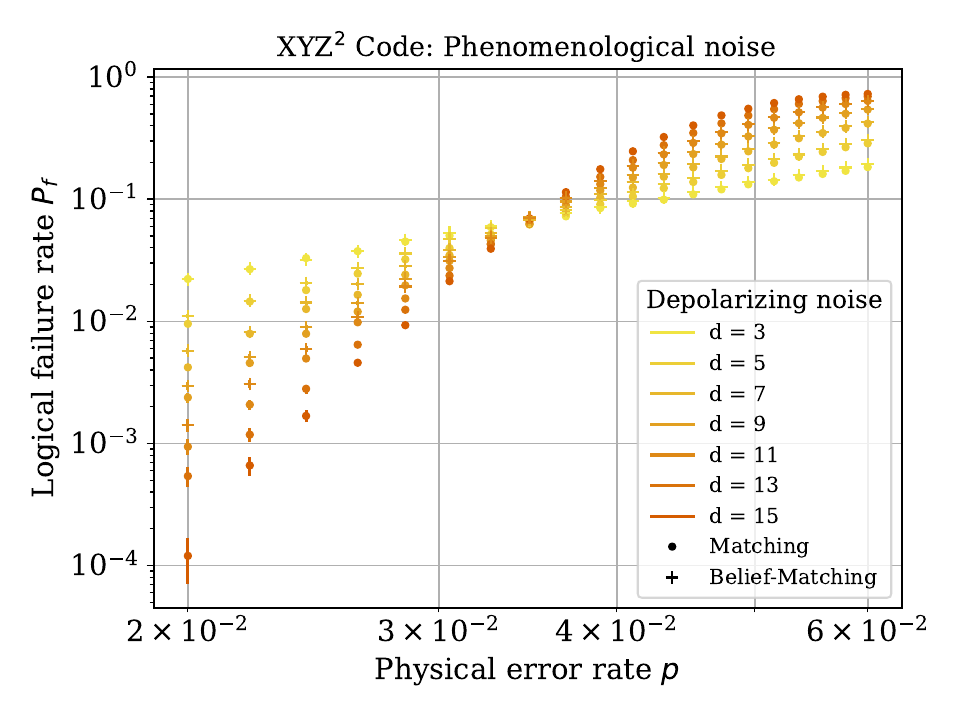}
\caption{Logical failure rate $P_f$ as a function of the physical error rate $p$ for different distances of the \xyz code, under a phenomenological depolarizing noise model ($\eta = 0.5$). Each data point is evaluated using $5 \times 10^4$ syndromes, decoded using a matching decoder (circular marker) or belief-matching decoder (plus marker) on the higher-level YZZY code. For depolarizing noise on the \xyz code, the noise model transforms to $Z$-biased noise on the YZZY code, which has a high threshold using a matching decoder. We observe a threshold of $3.42 \pm \SI{0.01}{\percent}$ for the matching decoder and $3.43 \pm \SI{0.01}{\percent}$ using a belief-matching decoder.}
\label{fig:phen-noise-matching-dep}
\end{figure}

\begin{figure}\centering
\includegraphics[width=\linewidth, trim={0.5cm 0.6cm 0.53cm 0.5cm},clip]{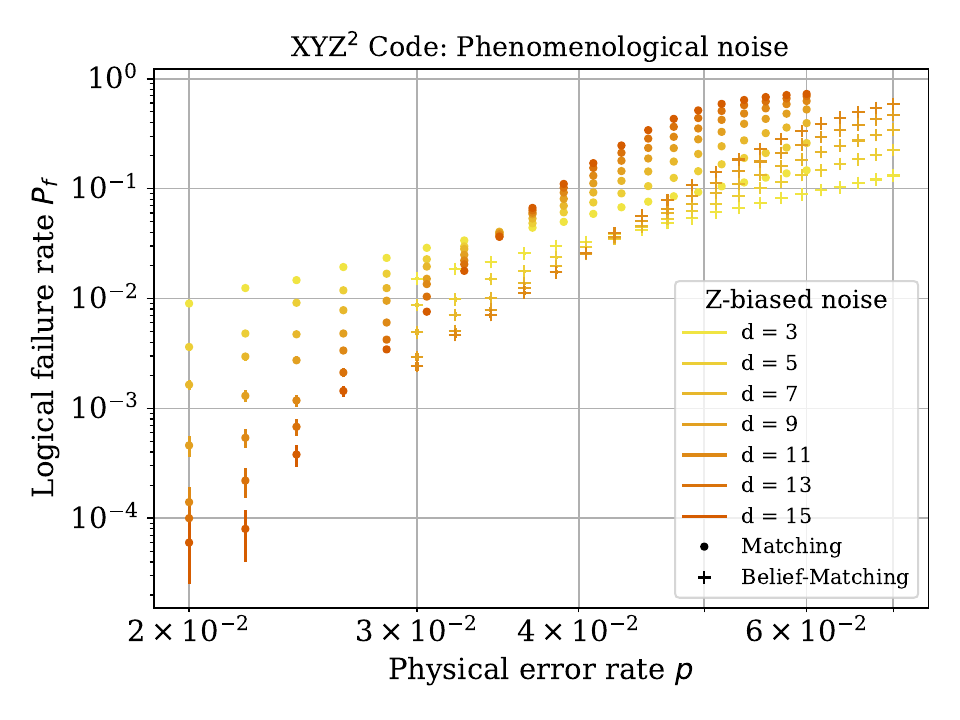}
\caption{Logical failure rate $P_f$ as a function of the physical error rate $p$ for different distances of the \xyz code, under a phenomenological $Z$-biased noise error model ($\eta = 10$). Each data point is evaluated using $5 \times 10^4$ syndromes, decoded using a matching decoder (circular marker) or a belief-matching decoder (plus marker) on the higher-level YZZY code. We observe a threshold of $3.49 \pm \SI{0.02}{\percent}$ for the matching decoder and $4.30 \pm \SI{0.02}{\percent}$ using a belief-matching decoder.}
\label{fig:phen-noise-matching-z10}
\end{figure}

Simulating phenomenological noise, we use the methods provided in \secref{subsec:pn-decoding} to decode the \xyz code both for depolarizing noise (see \figref{fig:phen-noise-matching-dep}) and for $Z$-biased noise with $\eta=10$ (see \figref{fig:phen-noise-matching-z10}). Similarly to the behavior for code-capacity noise we find that the belief-propagation-augmented decoder outperforms the matching decoder for biased noise, whereas the difference is marginal for code-capacity noise. In fact, as described in \appref{app:bm-vs-matching}, belief matching may perform slightly worse than matching for depolarizing noise models with imperfect syndrome measurements. Although tensor-network-based decoders are being developed for non-perfect stabilizers, they are limited to small code distances~\cite{PRXQuantum.5.040303} and have not been considered in this work. 

\section{Conclusion}
\label{sec:conclusion}

We developed a sequential decoding scheme for the \xyz code, which can be viewed as the concatenation of the YZZY code and a $\nkd{2}{1}{1}$ phase-error-detecting code. We implemented a sequential strategy for decoding the \xyz code by decoding the lower-level $\nkd{2}{1}{1}$ code optimally, sending the conditional logical error probabilities to the higher-level YZZY code, and decoding that further using MPS, matching, and belief-matching decoders.

We tested our decoding strategy by simulating both code-capacity and phenomenological noise, and provided algorithms for decoding the $XX$ links of the \xyz code in both cases. For the case of code-capacity noise, only one round of measurements is performed in the simulation and the links are decoded based on the optimal strategy (applying a $Z$ on an arbitrary qubit, and calculating conditional logical probabilities by brute force). For phenomenological noise, as the measurements are faulty, we simulated $d$ rounds of stabilizer measurements, for which the links are decoded separately first, and the plaquettes are decoded second by mapping to a YZZY code with phenomenological noise.

Decoding the upper-level YZZY code using information from the link decoding, using the MPS decoder, matching decoder, and belief-matching decoder, shows that the \xyz code decoded in a sequential fashion has high thresholds both for code-capacity as well as phenomenological noise. For code-capacity noise under a depolarizing noise model, the matching decoder gives a threshold of around \SI{18.3}{\percent}, which almost reaches the optimal threshold obtained using the MPS decoder of \SI{18.9}{\percent}. For $Z$-biased noise with $\eta = 10$, we observed that a matching decoder gives a threshold of \SI{18.4}{\percent}, which is not sufficient to reach the high threshold of \SI{28.6}{\percent} obtained by the optimal decoder. However, using a belief-propagation decoder in conjunction with the matching decoder increases the threshold to \SI{24.1}{\percent} and decreases the logical failure rate. 

Similarly, for a phenomenological noise model, for depolarizing noise, the matching decoder gives a threshold of \SI{3.4}{\percent}, which is on par with the belief-matching threshold of \SI{3.4}{\percent}. Whereas, for $Z$-biased noise with $\eta = 10$, belief-matching gives a higher threshold of \SI{4.3}{\percent}, compared to what we observed from using a matching decoder, \SI{3.5}{\percent}. While it was previously established \cite{Srivastava2022xyzhexagonal} that the \xyz code has high thresholds and low logical failure rates under code-capacity noise, comparable to the XZZX code, our work here demonstrates that high thresholds can be maintained by using efficient matching-based decoders as well, both for code-capacity noise and phenomenological noise models.

We showcased a method of decoding concatenated codes, using a particular instance of a non-CSS code, and demonstrated the advantages that can be achieved using belief-matching as a decoder for code-capacity and phenomenological noise. This work also exemplifies the advantages of using a higher-level code in a concatenated code scheme which is tailored to the noise model of the underlying logical qubits of the lower-level code.

The implementation of the sequential decoder for circuit-level noise, which would also require a fault-tolerant formulation of the stabilizer measurement circuits depending on the number and configuration of available ancilla qubits~\cite{hetenyi2024tailoring}, is left for future work.
Future research should also explore optimization of the placement of the erasure-removing $Z$ corrections on the links, by taking syndromes on the neighbouring plaquettes into consideration, instead of placing the corrections on the top qubit as is currently done. This strategy may improve the overall decoder performance when using a sub-optimal decoder for the upper-level surface code.

Finally, although in this work we considered $XX$ link stabilizers tailored to converting phase errors to erasure errors, it would also be interesting to explore the properties of the code using $-ZZ$ link-stabilizers, corresponding to a link code space of $|01\rangle$ and $|10\rangle$. With this formulation, the \xyz code is related to recently discussed dual-rail encoding~\cite{Chou2024, Teoh2023, Koottandavida2024, Levine2024}, which is tailored to qubit relaxation.

\begin{acknowledgments}

We thank Bence Het\'enyi for providing their decoder for the \xyz code and for assistance in running it. We thank James R.~Wootton and Benjamin J.~Brown for useful discussions. 
BS, AFK, and MG acknowledge financial support from the Knut and Alice Wallenberg Foundation through the Wallenberg Centre for Quantum Technology (WACQT). AFK also acknowledges support by the Swedish Research Council (grant number 2019-03696), the Swedish Foundation for Strategic Research (grants numbers FFL21-0279 and FUS21-0063), and the Horizon Europe programme HORIZON-CL4-2022-QUANTUM-01-SGA via the project 101113946 OpenSuperQPlus100. 
Computations were enabled by resources provided by the National Academic Infrastructure for Supercomputing in Sweden (NAISS), partially funded by the Swedish Research Council through grant agreement no.~2022-06725, and by Chalmers e-Commons through the Chalmers Centre for Computational Science and Engineering (C3SE).

\end{acknowledgments}

\bibliography{references}

\begin{thebibliography}{83}%
\makeatletter
\providecommand \@ifxundefined [1]{%
 \@ifx{#1\undefined}
}%
\providecommand \@ifnum [1]{%
 \ifnum #1\expandafter \@firstoftwo
 \else \expandafter \@secondoftwo
 \fi
}%
\providecommand \@ifx [1]{%
 \ifx #1\expandafter \@firstoftwo
 \else \expandafter \@secondoftwo
 \fi
}%
\providecommand \natexlab [1]{#1}%
\providecommand \enquote  [1]{``#1''}%
\providecommand \bibnamefont  [1]{#1}%
\providecommand \bibfnamefont [1]{#1}%
\providecommand \citenamefont [1]{#1}%
\providecommand \href@noop [0]{\@secondoftwo}%
\providecommand \href [0]{\begingroup \@sanitize@url \@href}%
\providecommand \@href[1]{\@@startlink{#1}\@@href}%
\providecommand \@@href[1]{\endgroup#1\@@endlink}%
\providecommand \@sanitize@url [0]{\catcode `\\12\catcode `\$12\catcode `\&12\catcode `\#12\catcode `\^12\catcode `\_12\catcode `\%12\relax}%
\providecommand \@@startlink[1]{}%
\providecommand \@@endlink[0]{}%
\providecommand \url  [0]{\begingroup\@sanitize@url \@url }%
\providecommand \@url [1]{\endgroup\@href {#1}{\urlprefix }}%
\providecommand \urlprefix  [0]{URL }%
\providecommand \Eprint [0]{\href }%
\providecommand \doibase [0]{https://doi.org/}%
\providecommand \selectlanguage [0]{\@gobble}%
\providecommand \bibinfo  [0]{\@secondoftwo}%
\providecommand \bibfield  [0]{\@secondoftwo}%
\providecommand \translation [1]{[#1]}%
\providecommand \BibitemOpen [0]{}%
\providecommand \bibitemStop [0]{}%
\providecommand \bibitemNoStop [0]{.\EOS\space}%
\providecommand \EOS [0]{\spacefactor3000\relax}%
\providecommand \BibitemShut  [1]{\csname bibitem#1\endcsname}%
\let\auto@bib@innerbib\@empty
\bibitem [{\citenamefont {Shor}(1994)}]{Shor1994Algorithms}%
  \BibitemOpen
  \bibfield  {author} {\bibinfo {author} {\bibfnamefont {P.}~\bibnamefont {Shor}},\ }\bibfield  {title} {\bibinfo {title} {Algorithms for quantum computation: discrete logarithms and factoring},\ }in\ \href {https://doi.org/10.1109/SFCS.1994.365700} {\emph {\bibinfo {booktitle} {Proceedings 35th Annual Symposium on Foundations of Computer Science}}}\ (\bibinfo {year} {1994})\ pp.\ \bibinfo {pages} {124--134}\BibitemShut {NoStop}%
\bibitem [{\citenamefont {Grover}(1997)}]{Grover1997}%
  \BibitemOpen
  \bibfield  {author} {\bibinfo {author} {\bibfnamefont {L.~K.}\ \bibnamefont {Grover}},\ }\bibfield  {title} {\bibinfo {title} {{Quantum Mechanics Helps in Searching for a Needle in a Haystack}},\ }\href {https://doi.org/10.1103/PhysRevLett.79.325} {\bibfield  {journal} {\bibinfo  {journal} {Physical Review Letters}\ }\textbf {\bibinfo {volume} {79}},\ \bibinfo {pages} {325} (\bibinfo {year} {1997})}\BibitemShut {NoStop}%
\bibitem [{\citenamefont {Jordan}()}]{quantumalgorithmzoo}%
  \BibitemOpen
  \bibfield  {author} {\bibinfo {author} {\bibfnamefont {S.}~\bibnamefont {Jordan}},\ }\href@noop {} {\bibinfo {title} {{Q}uantum {A}lgorithm {Z}oo --- quantumalgorithmzoo.org}},\ \bibinfo {howpublished} {\url{https://quantumalgorithmzoo.org/}}\BibitemShut {NoStop}%
\bibitem [{\citenamefont {Montanaro}(2016)}]{Montanaro2016}%
  \BibitemOpen
  \bibfield  {author} {\bibinfo {author} {\bibfnamefont {A.}~\bibnamefont {Montanaro}},\ }\bibfield  {title} {\bibinfo {title} {{Quantum algorithms: an overview}},\ }\href {https://doi.org/10.1038/npjqi.2015.23} {\bibfield  {journal} {\bibinfo  {journal} {npj Quantum Information}\ }\textbf {\bibinfo {volume} {2}},\ \bibinfo {pages} {15023} (\bibinfo {year} {2016})}\BibitemShut {NoStop}%
\bibitem [{\citenamefont {Dalzell}\ \emph {et~al.}(2023)\citenamefont {Dalzell}, \citenamefont {McArdle}, \citenamefont {Berta}, \citenamefont {Bienias}, \citenamefont {Chen}, \citenamefont {Gily{\'{e}}n}, \citenamefont {Hann}, \citenamefont {Kastoryano}, \citenamefont {Khabiboulline}, \citenamefont {Kubica}, \citenamefont {Salton}, \citenamefont {Wang},\ and\ \citenamefont {Brand{\~{a}}o}}]{Dalzell2023}%
  \BibitemOpen
  \bibfield  {author} {\bibinfo {author} {\bibfnamefont {A.~M.}\ \bibnamefont {Dalzell}}, \bibinfo {author} {\bibfnamefont {S.}~\bibnamefont {McArdle}}, \bibinfo {author} {\bibfnamefont {M.}~\bibnamefont {Berta}}, \bibinfo {author} {\bibfnamefont {P.}~\bibnamefont {Bienias}}, \bibinfo {author} {\bibfnamefont {C.-F.}\ \bibnamefont {Chen}}, \bibinfo {author} {\bibfnamefont {A.}~\bibnamefont {Gily{\'{e}}n}}, \bibinfo {author} {\bibfnamefont {C.~T.}\ \bibnamefont {Hann}}, \bibinfo {author} {\bibfnamefont {M.~J.}\ \bibnamefont {Kastoryano}}, \bibinfo {author} {\bibfnamefont {E.~T.}\ \bibnamefont {Khabiboulline}}, \bibinfo {author} {\bibfnamefont {A.}~\bibnamefont {Kubica}}, \bibinfo {author} {\bibfnamefont {G.}~\bibnamefont {Salton}}, \bibinfo {author} {\bibfnamefont {S.}~\bibnamefont {Wang}},\ and\ \bibinfo {author} {\bibfnamefont {F.~G. S.~L.}\ \bibnamefont {Brand{\~{a}}o}},\ }\href@noop {} {\bibinfo {title} {{Quantum algorithms: A survey of applications and end-to-end complexities}}} (\bibinfo {year} {2023}),\
  \Eprint {https://arxiv.org/abs/2310.03011} {arXiv:2310.03011} \BibitemShut {NoStop}%
\bibitem [{\citenamefont {Fowler}\ \emph {et~al.}(2012)\citenamefont {Fowler}, \citenamefont {Mariantoni}, \citenamefont {Martinis},\ and\ \citenamefont {Cleland}}]{Fowler2012}%
  \BibitemOpen
  \bibfield  {author} {\bibinfo {author} {\bibfnamefont {A.~G.}\ \bibnamefont {Fowler}}, \bibinfo {author} {\bibfnamefont {M.}~\bibnamefont {Mariantoni}}, \bibinfo {author} {\bibfnamefont {J.~M.}\ \bibnamefont {Martinis}},\ and\ \bibinfo {author} {\bibfnamefont {A.~N.}\ \bibnamefont {Cleland}},\ }\bibfield  {title} {\bibinfo {title} {Surface codes: Towards practical large-scale quantum computation},\ }\href {https://doi.org/10.1103/physreva.86.032324} {\bibfield  {journal} {\bibinfo  {journal} {Physical Review A}\ }\textbf {\bibinfo {volume} {86}},\ \bibinfo {pages} {032324} (\bibinfo {year} {2012})}\BibitemShut {NoStop}%
\bibitem [{\citenamefont {Preskill}(2018)}]{Preskill2018}%
  \BibitemOpen
  \bibfield  {author} {\bibinfo {author} {\bibfnamefont {J.}~\bibnamefont {Preskill}},\ }\bibfield  {title} {\bibinfo {title} {{Quantum Computing in the NISQ era and beyond}},\ }\href {https://doi.org/10.22331/q-2018-08-06-79} {\bibfield  {journal} {\bibinfo  {journal} {Quantum}\ }\textbf {\bibinfo {volume} {2}},\ \bibinfo {pages} {79} (\bibinfo {year} {2018})}\BibitemShut {NoStop}%
\bibitem [{\citenamefont {Kivlichan}\ \emph {et~al.}(2020)\citenamefont {Kivlichan}, \citenamefont {Gidney}, \citenamefont {Berry}, \citenamefont {Wiebe}, \citenamefont {McClean}, \citenamefont {Sun}, \citenamefont {Jiang}, \citenamefont {Rubin}, \citenamefont {Fowler}, \citenamefont {Aspuru-Guzik}, \citenamefont {Neven},\ and\ \citenamefont {Babbush}}]{Kivlichan2020}%
  \BibitemOpen
  \bibfield  {author} {\bibinfo {author} {\bibfnamefont {I.~D.}\ \bibnamefont {Kivlichan}}, \bibinfo {author} {\bibfnamefont {C.}~\bibnamefont {Gidney}}, \bibinfo {author} {\bibfnamefont {D.~W.}\ \bibnamefont {Berry}}, \bibinfo {author} {\bibfnamefont {N.}~\bibnamefont {Wiebe}}, \bibinfo {author} {\bibfnamefont {J.}~\bibnamefont {McClean}}, \bibinfo {author} {\bibfnamefont {W.}~\bibnamefont {Sun}}, \bibinfo {author} {\bibfnamefont {Z.}~\bibnamefont {Jiang}}, \bibinfo {author} {\bibfnamefont {N.}~\bibnamefont {Rubin}}, \bibinfo {author} {\bibfnamefont {A.}~\bibnamefont {Fowler}}, \bibinfo {author} {\bibfnamefont {A.}~\bibnamefont {Aspuru-Guzik}}, \bibinfo {author} {\bibfnamefont {H.}~\bibnamefont {Neven}},\ and\ \bibinfo {author} {\bibfnamefont {R.}~\bibnamefont {Babbush}},\ }\bibfield  {title} {\bibinfo {title} {{Improved Fault-Tolerant Quantum Simulation of Condensed-Phase Correlated Electrons via Trotterization}},\ }\href {https://doi.org/10.22331/q-2020-07-16-296} {\bibfield  {journal} {\bibinfo  {journal}
  {Quantum}\ }\textbf {\bibinfo {volume} {4}},\ \bibinfo {pages} {296} (\bibinfo {year} {2020})}\BibitemShut {NoStop}%
\bibitem [{\citenamefont {Campbell}(2021)}]{Campbell2021}%
  \BibitemOpen
  \bibfield  {author} {\bibinfo {author} {\bibfnamefont {E.~T.}\ \bibnamefont {Campbell}},\ }\bibfield  {title} {\bibinfo {title} {{Early fault-tolerant simulations of the Hubbard model}},\ }\href {https://doi.org/10.1088/2058-9565/ac3110} {\bibfield  {journal} {\bibinfo  {journal} {Quantum Science and Technology}\ }\textbf {\bibinfo {volume} {7}},\ \bibinfo {pages} {015007} (\bibinfo {year} {2021})}\BibitemShut {NoStop}%
\bibitem [{\citenamefont {Nielsen}\ and\ \citenamefont {Chuang}(2000)}]{Nielsen2000}%
  \BibitemOpen
  \bibfield  {author} {\bibinfo {author} {\bibfnamefont {M.~A.}\ \bibnamefont {Nielsen}}\ and\ \bibinfo {author} {\bibfnamefont {I.~L.}\ \bibnamefont {Chuang}},\ }\href@noop {} {\emph {\bibinfo {title} {{Quantum Computation and Quantum Information}}}}\ (\bibinfo  {publisher} {Cambridge University Press},\ \bibinfo {year} {2000})\BibitemShut {NoStop}%
\bibitem [{\citenamefont {Devitt}\ \emph {et~al.}(2013)\citenamefont {Devitt}, \citenamefont {Munro},\ and\ \citenamefont {Nemoto}}]{Devitt2013}%
  \BibitemOpen
  \bibfield  {author} {\bibinfo {author} {\bibfnamefont {S.~J.}\ \bibnamefont {Devitt}}, \bibinfo {author} {\bibfnamefont {W.~J.}\ \bibnamefont {Munro}},\ and\ \bibinfo {author} {\bibfnamefont {K.}~\bibnamefont {Nemoto}},\ }\bibfield  {title} {\bibinfo {title} {{Quantum error correction for beginners}},\ }\href {https://doi.org/10.1088/0034-4885/76/7/076001} {\bibfield  {journal} {\bibinfo  {journal} {Reports on Progress in Physics}\ }\textbf {\bibinfo {volume} {76}},\ \bibinfo {pages} {076001} (\bibinfo {year} {2013})}\BibitemShut {NoStop}%
\bibitem [{\citenamefont {Terhal}(2015)}]{Terhal2015}%
  \BibitemOpen
  \bibfield  {author} {\bibinfo {author} {\bibfnamefont {B.~M.}\ \bibnamefont {Terhal}},\ }\bibfield  {title} {\bibinfo {title} {{Quantum error correction for quantum memories}},\ }\href {https://doi.org/10.1103/RevModPhys.87.307} {\bibfield  {journal} {\bibinfo  {journal} {Reviews of Modern Physics}\ }\textbf {\bibinfo {volume} {87}},\ \bibinfo {pages} {307} (\bibinfo {year} {2015})}\BibitemShut {NoStop}%
\bibitem [{\citenamefont {Kitaev}(2003)}]{Kitaev2003}%
  \BibitemOpen
  \bibfield  {author} {\bibinfo {author} {\bibfnamefont {A.}~\bibnamefont {Kitaev}},\ }\bibfield  {title} {\bibinfo {title} {Fault-tolerant quantum computation by anyons},\ }\href {https://doi.org/10.1016/s0003-4916(02)00018-0} {\bibfield  {journal} {\bibinfo  {journal} {Annals of Physics}\ }\textbf {\bibinfo {volume} {303}},\ \bibinfo {pages} {2} (\bibinfo {year} {2003})}\BibitemShut {NoStop}%
\bibitem [{\citenamefont {Andersen}\ \emph {et~al.}(2020)\citenamefont {Andersen}, \citenamefont {Remm}, \citenamefont {Lazar}, \citenamefont {Krinner}, \citenamefont {Lacroix}, \citenamefont {Norris}, \citenamefont {Gabureac}, \citenamefont {Eichler},\ and\ \citenamefont {Wallraff}}]{Andersen2020}%
  \BibitemOpen
  \bibfield  {author} {\bibinfo {author} {\bibfnamefont {C.~K.}\ \bibnamefont {Andersen}}, \bibinfo {author} {\bibfnamefont {A.}~\bibnamefont {Remm}}, \bibinfo {author} {\bibfnamefont {S.}~\bibnamefont {Lazar}}, \bibinfo {author} {\bibfnamefont {S.}~\bibnamefont {Krinner}}, \bibinfo {author} {\bibfnamefont {N.}~\bibnamefont {Lacroix}}, \bibinfo {author} {\bibfnamefont {G.~J.}\ \bibnamefont {Norris}}, \bibinfo {author} {\bibfnamefont {M.}~\bibnamefont {Gabureac}}, \bibinfo {author} {\bibfnamefont {C.}~\bibnamefont {Eichler}},\ and\ \bibinfo {author} {\bibfnamefont {A.}~\bibnamefont {Wallraff}},\ }\bibfield  {title} {\bibinfo {title} {Repeated quantum error detection in a surface code},\ }\href {https://doi.org/10.1038/s41567-020-0920-y} {\bibfield  {journal} {\bibinfo  {journal} {Nature Physics}\ }\textbf {\bibinfo {volume} {16}},\ \bibinfo {pages} {875} (\bibinfo {year} {2020})}\BibitemShut {NoStop}%
\bibitem [{\citenamefont {Satzinger}\ \emph {et~al.}(2021)\citenamefont {Satzinger} \emph {et~al.}}]{Satzinger2021}%
  \BibitemOpen
  \bibfield  {author} {\bibinfo {author} {\bibfnamefont {K.~J.}\ \bibnamefont {Satzinger}} \emph {et~al.},\ }\bibfield  {title} {\bibinfo {title} {Realizing topologically ordered states on a quantum processor},\ }\href {https://doi.org/10.1126/science.abi8378} {\bibfield  {journal} {\bibinfo  {journal} {Science}\ }\textbf {\bibinfo {volume} {374}},\ \bibinfo {pages} {1237} (\bibinfo {year} {2021})}\BibitemShut {NoStop}%
\bibitem [{\citenamefont {Marques}\ \emph {et~al.}(2021)\citenamefont {Marques}, \citenamefont {Varbanov}, \citenamefont {Moreira}, \citenamefont {Ali}, \citenamefont {Muthusubramanian}, \citenamefont {Zachariadis}, \citenamefont {Battistel}, \citenamefont {Beekman}, \citenamefont {Haider}, \citenamefont {Vlothuizen}, \citenamefont {Bruno}, \citenamefont {Terhal},\ and\ \citenamefont {DiCarlo}}]{Marques2021}%
  \BibitemOpen
  \bibfield  {author} {\bibinfo {author} {\bibfnamefont {J.~F.}\ \bibnamefont {Marques}}, \bibinfo {author} {\bibfnamefont {B.~M.}\ \bibnamefont {Varbanov}}, \bibinfo {author} {\bibfnamefont {M.~S.}\ \bibnamefont {Moreira}}, \bibinfo {author} {\bibfnamefont {H.}~\bibnamefont {Ali}}, \bibinfo {author} {\bibfnamefont {N.}~\bibnamefont {Muthusubramanian}}, \bibinfo {author} {\bibfnamefont {C.}~\bibnamefont {Zachariadis}}, \bibinfo {author} {\bibfnamefont {F.}~\bibnamefont {Battistel}}, \bibinfo {author} {\bibfnamefont {M.}~\bibnamefont {Beekman}}, \bibinfo {author} {\bibfnamefont {N.}~\bibnamefont {Haider}}, \bibinfo {author} {\bibfnamefont {W.}~\bibnamefont {Vlothuizen}}, \bibinfo {author} {\bibfnamefont {A.}~\bibnamefont {Bruno}}, \bibinfo {author} {\bibfnamefont {B.~M.}\ \bibnamefont {Terhal}},\ and\ \bibinfo {author} {\bibfnamefont {L.}~\bibnamefont {DiCarlo}},\ }\bibfield  {title} {\bibinfo {title} {Logical-qubit operations in an error-detecting surface code},\ }\href
  {https://doi.org/10.1038/s41567-021-01423-9} {\bibfield  {journal} {\bibinfo  {journal} {Nature Physics}\ }\textbf {\bibinfo {volume} {18}},\ \bibinfo {pages} {80} (\bibinfo {year} {2021})}\BibitemShut {NoStop}%
\bibitem [{\citenamefont {Krinner}\ \emph {et~al.}(2022)\citenamefont {Krinner}, \citenamefont {Lacroix}, \citenamefont {Remm}, \citenamefont {Di~Paolo}, \citenamefont {Genois}, \citenamefont {Leroux}, \citenamefont {Hellings}, \citenamefont {Lazar}, \citenamefont {Swiadek}, \citenamefont {Herrmann}, \citenamefont {Norris}, \citenamefont {Andersen}, \citenamefont {M\"{u}ller}, \citenamefont {Blais}, \citenamefont {Eichler},\ and\ \citenamefont {Wallraff}}]{Krinner2022}%
  \BibitemOpen
  \bibfield  {author} {\bibinfo {author} {\bibfnamefont {S.}~\bibnamefont {Krinner}}, \bibinfo {author} {\bibfnamefont {N.}~\bibnamefont {Lacroix}}, \bibinfo {author} {\bibfnamefont {A.}~\bibnamefont {Remm}}, \bibinfo {author} {\bibfnamefont {A.}~\bibnamefont {Di~Paolo}}, \bibinfo {author} {\bibfnamefont {E.}~\bibnamefont {Genois}}, \bibinfo {author} {\bibfnamefont {C.}~\bibnamefont {Leroux}}, \bibinfo {author} {\bibfnamefont {C.}~\bibnamefont {Hellings}}, \bibinfo {author} {\bibfnamefont {S.}~\bibnamefont {Lazar}}, \bibinfo {author} {\bibfnamefont {F.}~\bibnamefont {Swiadek}}, \bibinfo {author} {\bibfnamefont {J.}~\bibnamefont {Herrmann}}, \bibinfo {author} {\bibfnamefont {G.~J.}\ \bibnamefont {Norris}}, \bibinfo {author} {\bibfnamefont {C.~K.}\ \bibnamefont {Andersen}}, \bibinfo {author} {\bibfnamefont {M.}~\bibnamefont {M\"{u}ller}}, \bibinfo {author} {\bibfnamefont {A.}~\bibnamefont {Blais}}, \bibinfo {author} {\bibfnamefont {C.}~\bibnamefont {Eichler}},\ and\ \bibinfo {author} {\bibfnamefont
  {A.}~\bibnamefont {Wallraff}},\ }\bibfield  {title} {\bibinfo {title} {Realizing repeated quantum error correction in a distance-three surface code},\ }\href {https://doi.org/10.1038/s41586-022-04566-8} {\bibfield  {journal} {\bibinfo  {journal} {Nature}\ }\textbf {\bibinfo {volume} {605}},\ \bibinfo {pages} {669} (\bibinfo {year} {2022})}\BibitemShut {NoStop}%
\bibitem [{\citenamefont {Acharya}\ \emph {et~al.}(2023)\citenamefont {Acharya} \emph {et~al.}}]{Acharya2023Suppressing}%
  \BibitemOpen
  \bibfield  {author} {\bibinfo {author} {\bibfnamefont {R.}~\bibnamefont {Acharya}} \emph {et~al.},\ }\bibfield  {title} {\bibinfo {title} {Suppressing quantum errors by scaling a surface code logical qubit},\ }\href {https://doi.org/10.1038/s41586-022-05434-1} {\bibfield  {journal} {\bibinfo  {journal} {Nature}\ }\textbf {\bibinfo {volume} {614}},\ \bibinfo {pages} {676} (\bibinfo {year} {2023})}\BibitemShut {NoStop}%
\bibitem [{\citenamefont {Acharya}\ \emph {et~al.}(2025)\citenamefont {Acharya} \emph {et~al.}}]{Acharya2025}%
  \BibitemOpen
  \bibfield  {author} {\bibinfo {author} {\bibfnamefont {R.}~\bibnamefont {Acharya}} \emph {et~al.},\ }\bibfield  {title} {\bibinfo {title} {Quantum error correction below the surface code threshold},\ }\href {https://doi.org/10.1038/s41586-024-08449-y} {\bibfield  {journal} {\bibinfo  {journal} {Nature}\ }\textbf {\bibinfo {volume} {638}},\ \bibinfo {pages} {920–926} (\bibinfo {year} {2025})}\BibitemShut {NoStop}%
\bibitem [{\citenamefont {Criger}\ and\ \citenamefont {Terhal}(2016)}]{Criger2016Noise}%
  \BibitemOpen
  \bibfield  {author} {\bibinfo {author} {\bibfnamefont {B.}~\bibnamefont {Criger}}\ and\ \bibinfo {author} {\bibfnamefont {B.}~\bibnamefont {Terhal}},\ }\bibfield  {title} {\bibinfo {title} {Noise thresholds for the [4,2,2]-concatenated toric code},\ }\href {https://doi.org/10.26421/qic16.15-16-1} {\bibfield  {journal} {\bibinfo  {journal} {Quantum Information and Computation}\ }\textbf {\bibinfo {volume} {16}},\ \bibinfo {pages} {1261} (\bibinfo {year} {2016})}\BibitemShut {NoStop}%
\bibitem [{\citenamefont {Yamasaki}\ and\ \citenamefont {Koashi}(2024)}]{Yamasaki2024Time}%
  \BibitemOpen
  \bibfield  {author} {\bibinfo {author} {\bibfnamefont {H.}~\bibnamefont {Yamasaki}}\ and\ \bibinfo {author} {\bibfnamefont {M.}~\bibnamefont {Koashi}},\ }\bibfield  {title} {\bibinfo {title} {{Time-Efficient Constant-Space-Overhead Fault-Tolerant Quantum Computation}},\ }\href {https://doi.org/10.1038/s41567-023-02325-8} {\bibfield  {journal} {\bibinfo  {journal} {Nature Physics}\ }\textbf {\bibinfo {volume} {20}},\ \bibinfo {pages} {247} (\bibinfo {year} {2024})}\BibitemShut {NoStop}%
\bibitem [{\citenamefont {Gidney}\ \emph {et~al.}(2023)\citenamefont {Gidney}, \citenamefont {Newman}, \citenamefont {Brooks},\ and\ \citenamefont {Jones}}]{Gidney2023Yoked}%
  \BibitemOpen
  \bibfield  {author} {\bibinfo {author} {\bibfnamefont {C.}~\bibnamefont {Gidney}}, \bibinfo {author} {\bibfnamefont {M.}~\bibnamefont {Newman}}, \bibinfo {author} {\bibfnamefont {P.}~\bibnamefont {Brooks}},\ and\ \bibinfo {author} {\bibfnamefont {C.}~\bibnamefont {Jones}},\ }\href@noop {} {\bibinfo {title} {Yoked surface codes}} (\bibinfo {year} {2023}),\ \Eprint {https://arxiv.org/abs/2312.04522} {arXiv:2312.04522} \BibitemShut {NoStop}%
\bibitem [{\citenamefont {Pattison}\ \emph {et~al.}(2023)\citenamefont {Pattison}, \citenamefont {Krishna},\ and\ \citenamefont {Preskill}}]{Pattison2023Hierarchical}%
  \BibitemOpen
  \bibfield  {author} {\bibinfo {author} {\bibfnamefont {C.~A.}\ \bibnamefont {Pattison}}, \bibinfo {author} {\bibfnamefont {A.}~\bibnamefont {Krishna}},\ and\ \bibinfo {author} {\bibfnamefont {J.}~\bibnamefont {Preskill}},\ }\href@noop {} {\bibinfo {title} {{Hierarchical memories: Simulating quantum LDPC codes with local gates}}} (\bibinfo {year} {2023}),\ \Eprint {https://arxiv.org/abs/2303.04798} {arXiv:2303.04798} \BibitemShut {NoStop}%
\bibitem [{\citenamefont {Yoshida}\ \emph {et~al.}(2024)\citenamefont {Yoshida}, \citenamefont {Tamiya},\ and\ \citenamefont {Yamasaki}}]{Satoshi2024Concatenate}%
  \BibitemOpen
  \bibfield  {author} {\bibinfo {author} {\bibfnamefont {S.}~\bibnamefont {Yoshida}}, \bibinfo {author} {\bibfnamefont {S.}~\bibnamefont {Tamiya}},\ and\ \bibinfo {author} {\bibfnamefont {H.}~\bibnamefont {Yamasaki}},\ }\href@noop {} {\bibinfo {title} {Concatenate codes, save qubits}} (\bibinfo {year} {2024}),\ \Eprint {https://arxiv.org/abs/2402.09606} {arXiv:2402.09606} \BibitemShut {NoStop}%
\bibitem [{\citenamefont {Vasmer}\ and\ \citenamefont {Kubica}(2022)}]{Vasmer2021Morphing}%
  \BibitemOpen
  \bibfield  {author} {\bibinfo {author} {\bibfnamefont {M.}~\bibnamefont {Vasmer}}\ and\ \bibinfo {author} {\bibfnamefont {A.}~\bibnamefont {Kubica}},\ }\bibfield  {title} {\bibinfo {title} {{Morphing Quantum Codes}},\ }\href {https://doi.org/10.1103/PRXQuantum.3.030319} {\bibfield  {journal} {\bibinfo  {journal} {PRX Quantum}\ }\textbf {\bibinfo {volume} {3}},\ \bibinfo {pages} {030319} (\bibinfo {year} {2022})}\BibitemShut {NoStop}%
\bibitem [{\citenamefont {Goto}(2024)}]{Goto2024Hypercube}%
  \BibitemOpen
  \bibfield  {author} {\bibinfo {author} {\bibfnamefont {H.}~\bibnamefont {Goto}},\ }\bibfield  {title} {\bibinfo {title} {{High-performance fault-tolerant quantum computing with many-hypercube codes}},\ }\href {https://doi.org/10.1126/sciadv.adp6388} {\bibfield  {journal} {\bibinfo  {journal} {Science Advances}\ }\textbf {\bibinfo {volume} {10}},\ \bibinfo {pages} {eadp6388} (\bibinfo {year} {2024})}\BibitemShut {NoStop}%
\bibitem [{\citenamefont {Li}\ \emph {et~al.}(2023)\citenamefont {Li}, \citenamefont {Kim},\ and\ \citenamefont {Hayden}}]{Li2023}%
  \BibitemOpen
  \bibfield  {author} {\bibinfo {author} {\bibfnamefont {Z.}~\bibnamefont {Li}}, \bibinfo {author} {\bibfnamefont {I.}~\bibnamefont {Kim}},\ and\ \bibinfo {author} {\bibfnamefont {P.}~\bibnamefont {Hayden}},\ }\bibfield  {title} {\bibinfo {title} {{Concatenation Schemes for Topological Fault-tolerant Quantum Error Correction}},\ }\href {https://doi.org/10.22331/q-2023-08-22-1089} {\bibfield  {journal} {\bibinfo  {journal} {Quantum}\ }\textbf {\bibinfo {volume} {7}},\ \bibinfo {pages} {1089} (\bibinfo {year} {2023})}\BibitemShut {NoStop}%
\bibitem [{\citenamefont {Meister}\ \emph {et~al.}(2024)\citenamefont {Meister}, \citenamefont {Pattison},\ and\ \citenamefont {Preskill}}]{Meister2024}%
  \BibitemOpen
  \bibfield  {author} {\bibinfo {author} {\bibfnamefont {N.}~\bibnamefont {Meister}}, \bibinfo {author} {\bibfnamefont {C.~A.}\ \bibnamefont {Pattison}},\ and\ \bibinfo {author} {\bibfnamefont {J.}~\bibnamefont {Preskill}},\ }\href@noop {} {\bibinfo {title} {Efficient soft-output decoders for the surface code}} (\bibinfo {year} {2024}),\ \Eprint {https://arxiv.org/abs/2405.07433} {arXiv:2405.07433} \BibitemShut {NoStop}%
\bibitem [{\citenamefont {Bonilla~Ataides}\ \emph {et~al.}(2021)\citenamefont {Bonilla~Ataides}, \citenamefont {Tuckett}, \citenamefont {Bartlett}, \citenamefont {Flammia},\ and\ \citenamefont {Brown}}]{Ataides2021XZZX}%
  \BibitemOpen
  \bibfield  {author} {\bibinfo {author} {\bibfnamefont {J.~P.}\ \bibnamefont {Bonilla~Ataides}}, \bibinfo {author} {\bibfnamefont {D.~K.}\ \bibnamefont {Tuckett}}, \bibinfo {author} {\bibfnamefont {S.~D.}\ \bibnamefont {Bartlett}}, \bibinfo {author} {\bibfnamefont {S.~T.}\ \bibnamefont {Flammia}},\ and\ \bibinfo {author} {\bibfnamefont {B.~J.}\ \bibnamefont {Brown}},\ }\bibfield  {title} {\bibinfo {title} {{The XZZX surface code}},\ }\href {https://doi.org/10.1038/s41467-021-22274-1} {\bibfield  {journal} {\bibinfo  {journal} {Nature Communications}\ }\textbf {\bibinfo {volume} {12}},\ \bibinfo {pages} {2172} (\bibinfo {year} {2021})}\BibitemShut {NoStop}%
\bibitem [{\citenamefont {Hastings}\ and\ \citenamefont {Haah}(2021)}]{Hastings2021}%
  \BibitemOpen
  \bibfield  {author} {\bibinfo {author} {\bibfnamefont {M.~B.}\ \bibnamefont {Hastings}}\ and\ \bibinfo {author} {\bibfnamefont {J.}~\bibnamefont {Haah}},\ }\bibfield  {title} {\bibinfo {title} {{Dynamically Generated Logical Qubits}},\ }\href {https://doi.org/10.22331/q-2021-10-19-564} {\bibfield  {journal} {\bibinfo  {journal} {Quantum}\ }\textbf {\bibinfo {volume} {5}},\ \bibinfo {pages} {564} (\bibinfo {year} {2021})}\BibitemShut {NoStop}%
\bibitem [{\citenamefont {Haah}\ and\ \citenamefont {Hastings}(2022)}]{Haah2022}%
  \BibitemOpen
  \bibfield  {author} {\bibinfo {author} {\bibfnamefont {J.}~\bibnamefont {Haah}}\ and\ \bibinfo {author} {\bibfnamefont {M.~B.}\ \bibnamefont {Hastings}},\ }\bibfield  {title} {\bibinfo {title} {{Boundaries for the Honeycomb Code}},\ }\href {https://doi.org/10.22331/q-2022-04-21-693} {\bibfield  {journal} {\bibinfo  {journal} {Quantum}\ }\textbf {\bibinfo {volume} {6}},\ \bibinfo {pages} {693} (\bibinfo {year} {2022})}\BibitemShut {NoStop}%
\bibitem [{\citenamefont {Wootton}(2015)}]{Wootton2015}%
  \BibitemOpen
  \bibfield  {author} {\bibinfo {author} {\bibfnamefont {J.~R.}\ \bibnamefont {Wootton}},\ }\bibfield  {title} {\bibinfo {title} {{A family of stabilizer codes for $D({{\mathbb{Z}}_{2}})$ anyons and Majorana modes}},\ }\href {https://doi.org/10.1088/1751-8113/48/21/215302} {\bibfield  {journal} {\bibinfo  {journal} {Journal of Physics A: Mathematical and Theoretical}\ }\textbf {\bibinfo {volume} {48}},\ \bibinfo {pages} {215302} (\bibinfo {year} {2015})}\BibitemShut {NoStop}%
\bibitem [{\citenamefont {Srivastava}\ \emph {et~al.}(2022)\citenamefont {Srivastava}, \citenamefont {Frisk~Kockum},\ and\ \citenamefont {Granath}}]{Srivastava2022xyzhexagonal}%
  \BibitemOpen
  \bibfield  {author} {\bibinfo {author} {\bibfnamefont {B.}~\bibnamefont {Srivastava}}, \bibinfo {author} {\bibfnamefont {A.}~\bibnamefont {Frisk~Kockum}},\ and\ \bibinfo {author} {\bibfnamefont {M.}~\bibnamefont {Granath}},\ }\bibfield  {title} {\bibinfo {title} {The {XYZ}{$^2$} hexagonal stabilizer code},\ }\href {https://doi.org/10.22331/q-2022-04-27-698} {\bibfield  {journal} {\bibinfo  {journal} {{Quantum}}\ }\textbf {\bibinfo {volume} {6}},\ \bibinfo {pages} {698} (\bibinfo {year} {2022})}\BibitemShut {NoStop}%
\bibitem [{\citenamefont {Huang}\ \emph {et~al.}(2023)\citenamefont {Huang}, \citenamefont {Pesah}, \citenamefont {Chubb}, \citenamefont {Vasmer},\ and\ \citenamefont {Dua}}]{Huang2023}%
  \BibitemOpen
  \bibfield  {author} {\bibinfo {author} {\bibfnamefont {E.}~\bibnamefont {Huang}}, \bibinfo {author} {\bibfnamefont {A.}~\bibnamefont {Pesah}}, \bibinfo {author} {\bibfnamefont {C.~T.}\ \bibnamefont {Chubb}}, \bibinfo {author} {\bibfnamefont {M.}~\bibnamefont {Vasmer}},\ and\ \bibinfo {author} {\bibfnamefont {A.}~\bibnamefont {Dua}},\ }\bibfield  {title} {\bibinfo {title} {{Tailoring Three-Dimensional Topological Codes for Biased Noise}},\ }\href {https://doi.org/10.1103/prxquantum.4.030338} {\bibfield  {journal} {\bibinfo  {journal} {PRX Quantum}\ }\textbf {\bibinfo {volume} {4}},\ \bibinfo {pages} {030338} (\bibinfo {year} {2023})}\BibitemShut {NoStop}%
\bibitem [{\citenamefont {Wootton}(2022)}]{Wootton2022}%
  \BibitemOpen
  \bibfield  {author} {\bibinfo {author} {\bibfnamefont {J.~R.}\ \bibnamefont {Wootton}},\ }\bibfield  {title} {\bibinfo {title} {Hexagonal matching codes with two-body measurements},\ }\href {https://doi.org/10.1088/1751-8121/ac7a75} {\bibfield  {journal} {\bibinfo  {journal} {Journal of Physics A: Mathematical and Theoretical}\ }\textbf {\bibinfo {volume} {55}},\ \bibinfo {pages} {295302} (\bibinfo {year} {2022})}\BibitemShut {NoStop}%
\bibitem [{\citenamefont {Tuckett}\ \emph {et~al.}(2018)\citenamefont {Tuckett}, \citenamefont {Bartlett},\ and\ \citenamefont {Flammia}}]{Tuckett2018}%
  \BibitemOpen
  \bibfield  {author} {\bibinfo {author} {\bibfnamefont {D.~K.}\ \bibnamefont {Tuckett}}, \bibinfo {author} {\bibfnamefont {S.~D.}\ \bibnamefont {Bartlett}},\ and\ \bibinfo {author} {\bibfnamefont {S.~T.}\ \bibnamefont {Flammia}},\ }\bibfield  {title} {\bibinfo {title} {{Ultrahigh Error Threshold for Surface Codes with Biased Noise}},\ }\href {https://doi.org/10.1103/physrevlett.120.050505} {\bibfield  {journal} {\bibinfo  {journal} {Physical Review Letters}\ }\textbf {\bibinfo {volume} {120}},\ \bibinfo {pages} {050505} (\bibinfo {year} {2018})}\BibitemShut {NoStop}%
\bibitem [{\citenamefont {Tuckett}\ \emph {et~al.}(2019)\citenamefont {Tuckett}, \citenamefont {Darmawan}, \citenamefont {Chubb}, \citenamefont {Bravyi}, \citenamefont {Bartlett},\ and\ \citenamefont {Flammia}}]{Tuckett2019}%
  \BibitemOpen
  \bibfield  {author} {\bibinfo {author} {\bibfnamefont {D.~K.}\ \bibnamefont {Tuckett}}, \bibinfo {author} {\bibfnamefont {A.~S.}\ \bibnamefont {Darmawan}}, \bibinfo {author} {\bibfnamefont {C.~T.}\ \bibnamefont {Chubb}}, \bibinfo {author} {\bibfnamefont {S.}~\bibnamefont {Bravyi}}, \bibinfo {author} {\bibfnamefont {S.~D.}\ \bibnamefont {Bartlett}},\ and\ \bibinfo {author} {\bibfnamefont {S.~T.}\ \bibnamefont {Flammia}},\ }\bibfield  {title} {\bibinfo {title} {{Tailoring Surface Codes for Highly Biased Noise}},\ }\href {https://doi.org/10.1103/physrevx.9.041031} {\bibfield  {journal} {\bibinfo  {journal} {Physical Review X}\ }\textbf {\bibinfo {volume} {9}},\ \bibinfo {pages} {041031} (\bibinfo {year} {2019})}\BibitemShut {NoStop}%
\bibitem [{\citenamefont {Tuckett}\ \emph {et~al.}(2020)\citenamefont {Tuckett}, \citenamefont {Bartlett}, \citenamefont {Flammia},\ and\ \citenamefont {Brown}}]{Tuckett2020}%
  \BibitemOpen
  \bibfield  {author} {\bibinfo {author} {\bibfnamefont {D.~K.}\ \bibnamefont {Tuckett}}, \bibinfo {author} {\bibfnamefont {S.~D.}\ \bibnamefont {Bartlett}}, \bibinfo {author} {\bibfnamefont {S.~T.}\ \bibnamefont {Flammia}},\ and\ \bibinfo {author} {\bibfnamefont {B.~J.}\ \bibnamefont {Brown}},\ }\bibfield  {title} {\bibinfo {title} {{Fault-Tolerant Thresholds for the Surface Code in Excess of 5\% Under Biased Noise}},\ }\href {https://doi.org/10.1103/physrevlett.124.130501} {\bibfield  {journal} {\bibinfo  {journal} {Physical Review Letters}\ }\textbf {\bibinfo {volume} {124}},\ \bibinfo {pages} {130501} (\bibinfo {year} {2020})}\BibitemShut {NoStop}%
\bibitem [{\citenamefont {Dua}\ \emph {et~al.}(2024)\citenamefont {Dua}, \citenamefont {Kubica}, \citenamefont {Jiang}, \citenamefont {Flammia},\ and\ \citenamefont {Gullans}}]{Dua2024}%
  \BibitemOpen
  \bibfield  {author} {\bibinfo {author} {\bibfnamefont {A.}~\bibnamefont {Dua}}, \bibinfo {author} {\bibfnamefont {A.}~\bibnamefont {Kubica}}, \bibinfo {author} {\bibfnamefont {L.}~\bibnamefont {Jiang}}, \bibinfo {author} {\bibfnamefont {S.~T.}\ \bibnamefont {Flammia}},\ and\ \bibinfo {author} {\bibfnamefont {M.~J.}\ \bibnamefont {Gullans}},\ }\bibfield  {title} {\bibinfo {title} {{Clifford-Deformed Surface Codes}},\ }\href {https://doi.org/10.1103/prxquantum.5.010347} {\bibfield  {journal} {\bibinfo  {journal} {PRX Quantum}\ }\textbf {\bibinfo {volume} {5}},\ \bibinfo {pages} {010347} (\bibinfo {year} {2024})}\BibitemShut {NoStop}%
\bibitem [{\citenamefont {Xiao}\ \emph {et~al.}(2024)\citenamefont {Xiao}, \citenamefont {Srivastava},\ and\ \citenamefont {Granath}}]{Xiao2024Exact}%
  \BibitemOpen
  \bibfield  {author} {\bibinfo {author} {\bibfnamefont {Y.}~\bibnamefont {Xiao}}, \bibinfo {author} {\bibfnamefont {B.}~\bibnamefont {Srivastava}},\ and\ \bibinfo {author} {\bibfnamefont {M.}~\bibnamefont {Granath}},\ }\bibfield  {title} {\bibinfo {title} {Exact results on finite size corrections for surface codes tailored to biased noise},\ }\href {https://doi.org/10.22331/q-2024-09-11-1468} {\bibfield  {journal} {\bibinfo  {journal} {Quantum}\ }\textbf {\bibinfo {volume} {8}},\ \bibinfo {pages} {1468} (\bibinfo {year} {2024})}\BibitemShut {NoStop}%
\bibitem [{\citenamefont {Poulin}(2006)}]{Poulin2006Optimal}%
  \BibitemOpen
  \bibfield  {author} {\bibinfo {author} {\bibfnamefont {D.}~\bibnamefont {Poulin}},\ }\bibfield  {title} {\bibinfo {title} {Optimal and efficient decoding of concatenated quantum block codes},\ }\href {https://doi.org/10.1103/PhysRevA.74.052333} {\bibfield  {journal} {\bibinfo  {journal} {Physical Review A}\ }\textbf {\bibinfo {volume} {74}},\ \bibinfo {pages} {052333} (\bibinfo {year} {2006})}\BibitemShut {NoStop}%
\bibitem [{\citenamefont {Higgott}\ and\ \citenamefont {Gidney}(2022)}]{pymatchingv2}%
  \BibitemOpen
  \bibfield  {author} {\bibinfo {author} {\bibfnamefont {O.}~\bibnamefont {Higgott}}\ and\ \bibinfo {author} {\bibfnamefont {C.}~\bibnamefont {Gidney}},\ }\href@noop {} {\bibinfo {title} {Pymatching v2}},\ \bibinfo {howpublished} {\url{https://github.com/oscarhiggott/PyMatching}} (\bibinfo {year} {2022})\BibitemShut {NoStop}%
\bibitem [{\citenamefont {Criger}\ and\ \citenamefont {Ashraf}(2018)}]{Criger2018}%
  \BibitemOpen
  \bibfield  {author} {\bibinfo {author} {\bibfnamefont {B.}~\bibnamefont {Criger}}\ and\ \bibinfo {author} {\bibfnamefont {I.}~\bibnamefont {Ashraf}},\ }\bibfield  {title} {\bibinfo {title} {{Multi-path Summation for Decoding 2D Topological Codes}},\ }\href {https://doi.org/10.22331/q-2018-10-19-102} {\bibfield  {journal} {\bibinfo  {journal} {Quantum}\ }\textbf {\bibinfo {volume} {2}},\ \bibinfo {pages} {102} (\bibinfo {year} {2018})}\BibitemShut {NoStop}%
\bibitem [{\citenamefont {Higgott}\ \emph {et~al.}(2023)\citenamefont {Higgott}, \citenamefont {Bohdanowicz}, \citenamefont {Kubica}, \citenamefont {Flammia},\ and\ \citenamefont {Campbell}}]{Higgott2023ImprovedDecoding}%
  \BibitemOpen
  \bibfield  {author} {\bibinfo {author} {\bibfnamefont {O.}~\bibnamefont {Higgott}}, \bibinfo {author} {\bibfnamefont {T.~C.}\ \bibnamefont {Bohdanowicz}}, \bibinfo {author} {\bibfnamefont {A.}~\bibnamefont {Kubica}}, \bibinfo {author} {\bibfnamefont {S.~T.}\ \bibnamefont {Flammia}},\ and\ \bibinfo {author} {\bibfnamefont {E.~T.}\ \bibnamefont {Campbell}},\ }\bibfield  {title} {\bibinfo {title} {{Improved Decoding of Circuit Noise and Fragile Boundaries of Tailored Surface Codes}},\ }\href {https://doi.org/10.1103/PhysRevX.13.031007} {\bibfield  {journal} {\bibinfo  {journal} {Physical Review X}\ }\textbf {\bibinfo {volume} {13}},\ \bibinfo {pages} {031007} (\bibinfo {year} {2023})}\BibitemShut {NoStop}%
\bibitem [{\citenamefont {Bravyi}\ \emph {et~al.}(2014{\natexlab{a}})\citenamefont {Bravyi}, \citenamefont {Suchara},\ and\ \citenamefont {Vargo}}]{Bravyi2014EfficientCode}%
  \BibitemOpen
  \bibfield  {author} {\bibinfo {author} {\bibfnamefont {S.}~\bibnamefont {Bravyi}}, \bibinfo {author} {\bibfnamefont {M.}~\bibnamefont {Suchara}},\ and\ \bibinfo {author} {\bibfnamefont {A.}~\bibnamefont {Vargo}},\ }\bibfield  {title} {\bibinfo {title} {{Efficient algorithms for maximum likelihood decoding in the surface code}},\ }\href {https://doi.org/10.1103/PhysRevA.90.032326} {\bibfield  {journal} {\bibinfo  {journal} {Physical Review A}\ }\textbf {\bibinfo {volume} {90}},\ \bibinfo {pages} {032326} (\bibinfo {year} {2014}{\natexlab{a}})}\BibitemShut {NoStop}%
\bibitem [{\citenamefont {Tuckett}(2020)}]{qecsim}%
  \BibitemOpen
  \bibfield  {author} {\bibinfo {author} {\bibfnamefont {D.~K.}\ \bibnamefont {Tuckett}},\ }\emph {\bibinfo {title} {Tailoring surface codes: Improvements in quantum error correction with biased noise}},\ \href {https://doi.org/10.25910/x8xw-9077} {Ph.D. thesis},\ \bibinfo  {school} {University of Sydney} (\bibinfo {year} {2020}),\ \bibinfo {note} {(qecsim: \url{https://github.com/qecsim/qecsim})}\BibitemShut {NoStop}%
\bibitem [{\citenamefont {Het\'enyi}\ and\ \citenamefont {Wootton}(2024)}]{hetenyi2024tailoring}%
  \BibitemOpen
  \bibfield  {author} {\bibinfo {author} {\bibfnamefont {B.}~\bibnamefont {Het\'enyi}}\ and\ \bibinfo {author} {\bibfnamefont {J.~R.}\ \bibnamefont {Wootton}},\ }\bibfield  {title} {\bibinfo {title} {Tailoring quantum error correction to spin qubits},\ }\href {https://doi.org/10.1103/PhysRevA.109.032433} {\bibfield  {journal} {\bibinfo  {journal} {Physical Review A}\ }\textbf {\bibinfo {volume} {109}},\ \bibinfo {pages} {032433} (\bibinfo {year} {2024})}\BibitemShut {NoStop}%
\bibitem [{\citenamefont {Gottesman}(1997)}]{Gottesman1997Stabilizer}%
  \BibitemOpen
  \bibfield  {author} {\bibinfo {author} {\bibfnamefont {D.}~\bibnamefont {Gottesman}},\ }\emph {\bibinfo {title} {{Stabilizer Codes and Quantum Error Correction}}},\ \href {https://doi.org/10.48550/ARXIV.QUANT-PH/9705052} {Ph.D. thesis},\ \bibinfo  {school} {California Institute of Technology} (\bibinfo {year} {1997})\BibitemShut {NoStop}%
\bibitem [{\citenamefont {Gottesman}(2009)}]{Gottesman2009Introduction}%
  \BibitemOpen
  \bibfield  {author} {\bibinfo {author} {\bibfnamefont {D.}~\bibnamefont {Gottesman}},\ }\href@noop {} {\bibinfo {title} {{An Introduction to Quantum Error Correction and Fault-Tolerant Quantum Computation}}} (\bibinfo {year} {2009}),\ \Eprint {https://arxiv.org/abs/0904.2557} {arXiv:0904.2557} \BibitemShut {NoStop}%
\bibitem [{\citenamefont {deMarti iOlius}\ \emph {et~al.}(2024)\citenamefont {deMarti iOlius}, \citenamefont {Fuentes}, \citenamefont {Orús}, \citenamefont {Crespo},\ and\ \citenamefont {Etxezarreta~Martinez}}]{deMarti_iOlius_2024}%
  \BibitemOpen
  \bibfield  {author} {\bibinfo {author} {\bibfnamefont {A.}~\bibnamefont {deMarti iOlius}}, \bibinfo {author} {\bibfnamefont {P.}~\bibnamefont {Fuentes}}, \bibinfo {author} {\bibfnamefont {R.}~\bibnamefont {Orús}}, \bibinfo {author} {\bibfnamefont {P.~M.}\ \bibnamefont {Crespo}},\ and\ \bibinfo {author} {\bibfnamefont {J.}~\bibnamefont {Etxezarreta~Martinez}},\ }\bibfield  {title} {\bibinfo {title} {Decoding algorithms for surface codes},\ }\href {https://doi.org/10.22331/q-2024-10-10-1498} {\bibfield  {journal} {\bibinfo  {journal} {Quantum}\ }\textbf {\bibinfo {volume} {8}},\ \bibinfo {pages} {1498} (\bibinfo {year} {2024})}\BibitemShut {NoStop}%
\bibitem [{\citenamefont {Dennis}\ \emph {et~al.}(2002)\citenamefont {Dennis}, \citenamefont {Kitaev}, \citenamefont {Landahl},\ and\ \citenamefont {Preskill}}]{Dennis2002}%
  \BibitemOpen
  \bibfield  {author} {\bibinfo {author} {\bibfnamefont {E.}~\bibnamefont {Dennis}}, \bibinfo {author} {\bibfnamefont {A.}~\bibnamefont {Kitaev}}, \bibinfo {author} {\bibfnamefont {A.}~\bibnamefont {Landahl}},\ and\ \bibinfo {author} {\bibfnamefont {J.}~\bibnamefont {Preskill}},\ }\bibfield  {title} {\bibinfo {title} {Topological quantum memory},\ }\href {https://doi.org/10.1063/1.1499754} {\bibfield  {journal} {\bibinfo  {journal} {Journal of Mathematical Physics}\ }\textbf {\bibinfo {volume} {43}},\ \bibinfo {pages} {4452} (\bibinfo {year} {2002})}\BibitemShut {NoStop}%
\bibitem [{\citenamefont {Wang}\ \emph {et~al.}(2010)\citenamefont {Wang}, \citenamefont {Fowler}, \citenamefont {Stephens},\ and\ \citenamefont {Hollenberg}}]{wang2009threshold}%
  \BibitemOpen
  \bibfield  {author} {\bibinfo {author} {\bibfnamefont {D.~S.}\ \bibnamefont {Wang}}, \bibinfo {author} {\bibfnamefont {A.~G.}\ \bibnamefont {Fowler}}, \bibinfo {author} {\bibfnamefont {A.~M.}\ \bibnamefont {Stephens}},\ and\ \bibinfo {author} {\bibfnamefont {L.~C.~L.}\ \bibnamefont {Hollenberg}},\ }\bibfield  {title} {\bibinfo {title} {{Threshold Error Rates for the Toric and Planar Codes}},\ }\href {https://dl.acm.org/doi/abs/10.5555/2011362.2011368} {\bibfield  {journal} {\bibinfo  {journal} {Quantum Information \& Computation}\ }\textbf {\bibinfo {volume} {10}},\ \bibinfo {pages} {456} (\bibinfo {year} {2010})}\BibitemShut {NoStop}%
\bibitem [{\citenamefont {Poulin}\ and\ \citenamefont {Chung}(2008)}]{Poulin2008iterative}%
  \BibitemOpen
  \bibfield  {author} {\bibinfo {author} {\bibfnamefont {D.}~\bibnamefont {Poulin}}\ and\ \bibinfo {author} {\bibfnamefont {Y.}~\bibnamefont {Chung}},\ }\bibfield  {title} {\bibinfo {title} {{On the Iterative Decoding of Sparse Quantum Codes}},\ }\href {https://dl.acm.org/doi/10.5555/2016985.2016993} {\bibfield  {journal} {\bibinfo  {journal} {Quantum Information \& Computation}\ }\textbf {\bibinfo {volume} {8}},\ \bibinfo {pages} {987–1000} (\bibinfo {year} {2008})}\BibitemShut {NoStop}%
\bibitem [{\citenamefont {Roffe}\ \emph {et~al.}(2020)\citenamefont {Roffe}, \citenamefont {White}, \citenamefont {Burton},\ and\ \citenamefont {Campbell}}]{Roffe2020}%
  \BibitemOpen
  \bibfield  {author} {\bibinfo {author} {\bibfnamefont {J.}~\bibnamefont {Roffe}}, \bibinfo {author} {\bibfnamefont {D.~R.}\ \bibnamefont {White}}, \bibinfo {author} {\bibfnamefont {S.}~\bibnamefont {Burton}},\ and\ \bibinfo {author} {\bibfnamefont {E.}~\bibnamefont {Campbell}},\ }\bibfield  {title} {\bibinfo {title} {Decoding across the quantum low-density parity-check code landscape},\ }\href {https://doi.org/10.1103/physrevresearch.2.043423} {\bibfield  {journal} {\bibinfo  {journal} {Physical Review Research}\ }\textbf {\bibinfo {volume} {2}},\ \bibinfo {pages} {043423} (\bibinfo {year} {2020})}\BibitemShut {NoStop}%
\bibitem [{\citenamefont {Wootton}\ and\ \citenamefont {Loss}(2012)}]{Wootton2012}%
  \BibitemOpen
  \bibfield  {author} {\bibinfo {author} {\bibfnamefont {J.~R.}\ \bibnamefont {Wootton}}\ and\ \bibinfo {author} {\bibfnamefont {D.}~\bibnamefont {Loss}},\ }\bibfield  {title} {\bibinfo {title} {{High Threshold Error Correction for the Surface Code}},\ }\href {https://doi.org/10.1103/physrevlett.109.160503} {\bibfield  {journal} {\bibinfo  {journal} {Physical Review Letters}\ }\textbf {\bibinfo {volume} {109}},\ \bibinfo {pages} {160503} (\bibinfo {year} {2012})}\BibitemShut {NoStop}%
\bibitem [{\citenamefont {Bravyi}\ \emph {et~al.}(2014{\natexlab{b}})\citenamefont {Bravyi}, \citenamefont {Suchara},\ and\ \citenamefont {Vargo}}]{Bravyi2014}%
  \BibitemOpen
  \bibfield  {author} {\bibinfo {author} {\bibfnamefont {S.}~\bibnamefont {Bravyi}}, \bibinfo {author} {\bibfnamefont {M.}~\bibnamefont {Suchara}},\ and\ \bibinfo {author} {\bibfnamefont {A.}~\bibnamefont {Vargo}},\ }\bibfield  {title} {\bibinfo {title} {Efficient algorithms for maximum likelihood decoding in the surface code},\ }\href {https://doi.org/10.1103/physreva.90.032326} {\bibfield  {journal} {\bibinfo  {journal} {Physical Review A}\ }\textbf {\bibinfo {volume} {90}},\ \bibinfo {pages} {032326} (\bibinfo {year} {2014}{\natexlab{b}})}\BibitemShut {NoStop}%
\bibitem [{\citenamefont {Hutter}\ \emph {et~al.}(2014)\citenamefont {Hutter}, \citenamefont {Wootton},\ and\ \citenamefont {Loss}}]{Hutter2014}%
  \BibitemOpen
  \bibfield  {author} {\bibinfo {author} {\bibfnamefont {A.}~\bibnamefont {Hutter}}, \bibinfo {author} {\bibfnamefont {J.~R.}\ \bibnamefont {Wootton}},\ and\ \bibinfo {author} {\bibfnamefont {D.}~\bibnamefont {Loss}},\ }\bibfield  {title} {\bibinfo {title} {{Efficient Markov chain Monte Carlo algorithm for the surface code}},\ }\href {https://doi.org/10.1103/physreva.89.022326} {\bibfield  {journal} {\bibinfo  {journal} {Physical Review A}\ }\textbf {\bibinfo {volume} {89}},\ \bibinfo {pages} {022326} (\bibinfo {year} {2014})}\BibitemShut {NoStop}%
\bibitem [{\citenamefont {Hammar}\ \emph {et~al.}(2022)\citenamefont {Hammar}, \citenamefont {Orekhov}, \citenamefont {Hybelius}, \citenamefont {Wisakanto}, \citenamefont {Srivastava}, \citenamefont {Kockum},\ and\ \citenamefont {Granath}}]{Hammar2022}%
  \BibitemOpen
  \bibfield  {author} {\bibinfo {author} {\bibfnamefont {K.}~\bibnamefont {Hammar}}, \bibinfo {author} {\bibfnamefont {A.}~\bibnamefont {Orekhov}}, \bibinfo {author} {\bibfnamefont {P.~W.}\ \bibnamefont {Hybelius}}, \bibinfo {author} {\bibfnamefont {A.~K.}\ \bibnamefont {Wisakanto}}, \bibinfo {author} {\bibfnamefont {B.}~\bibnamefont {Srivastava}}, \bibinfo {author} {\bibfnamefont {A.~F.}\ \bibnamefont {Kockum}},\ and\ \bibinfo {author} {\bibfnamefont {M.}~\bibnamefont {Granath}},\ }\bibfield  {title} {\bibinfo {title} {Error-rate-agnostic decoding of topological stabilizer codes},\ }\href {https://doi.org/10.1103/physreva.105.042616} {\bibfield  {journal} {\bibinfo  {journal} {Physical Review A}\ }\textbf {\bibinfo {volume} {105}},\ \bibinfo {pages} {042616} (\bibinfo {year} {2022})}\BibitemShut {NoStop}%
\bibitem [{\citenamefont {Chubb}(2021)}]{chubb2021generaltensornetworkdecoding}%
  \BibitemOpen
  \bibfield  {author} {\bibinfo {author} {\bibfnamefont {C.~T.}\ \bibnamefont {Chubb}},\ }\href@noop {} {\bibinfo {title} {{General tensor network decoding of 2D Pauli codes}}} (\bibinfo {year} {2021}),\ \Eprint {https://arxiv.org/abs/2101.04125} {arXiv:2101.04125} \BibitemShut {NoStop}%
\bibitem [{\citenamefont {Lange}\ \emph {et~al.}(2023)\citenamefont {Lange}, \citenamefont {Havstr\"{o}m}, \citenamefont {Srivastava}, \citenamefont {Bergentall}, \citenamefont {Hammar}, \citenamefont {Heuts}, \citenamefont {van Nieuwenburg},\ and\ \citenamefont {Granath}}]{Lange2023data}%
  \BibitemOpen
  \bibfield  {author} {\bibinfo {author} {\bibfnamefont {M.}~\bibnamefont {Lange}}, \bibinfo {author} {\bibfnamefont {P.}~\bibnamefont {Havstr\"{o}m}}, \bibinfo {author} {\bibfnamefont {B.}~\bibnamefont {Srivastava}}, \bibinfo {author} {\bibfnamefont {V.}~\bibnamefont {Bergentall}}, \bibinfo {author} {\bibfnamefont {K.}~\bibnamefont {Hammar}}, \bibinfo {author} {\bibfnamefont {O.}~\bibnamefont {Heuts}}, \bibinfo {author} {\bibfnamefont {E.}~\bibnamefont {van Nieuwenburg}},\ and\ \bibinfo {author} {\bibfnamefont {M.}~\bibnamefont {Granath}},\ }\href@noop {} {\bibinfo {title} {Data-driven decoding of quantum error correcting codes using graph neural networks}} (\bibinfo {year} {2023}),\ \Eprint {https://arxiv.org/abs/2307.01241} {arXiv:2307.01241} \BibitemShut {NoStop}%
\bibitem [{\citenamefont {Varbanov}\ \emph {et~al.}(2023)\citenamefont {Varbanov}, \citenamefont {Serra-Peralta}, \citenamefont {Byfield},\ and\ \citenamefont {Terhal}}]{Varbanov2023Neural}%
  \BibitemOpen
  \bibfield  {author} {\bibinfo {author} {\bibfnamefont {B.~M.}\ \bibnamefont {Varbanov}}, \bibinfo {author} {\bibfnamefont {M.}~\bibnamefont {Serra-Peralta}}, \bibinfo {author} {\bibfnamefont {D.}~\bibnamefont {Byfield}},\ and\ \bibinfo {author} {\bibfnamefont {B.~M.}\ \bibnamefont {Terhal}},\ }\href@noop {} {\bibinfo {title} {Neural network decoder for near-term surface-code experiments}} (\bibinfo {year} {2023}),\ \Eprint {https://arxiv.org/abs/2307.03280} {arXiv:2307.03280} \BibitemShut {NoStop}%
\bibitem [{\citenamefont {Bausch}\ \emph {et~al.}(2024)\citenamefont {Bausch}, \citenamefont {Senior}, \citenamefont {Heras}, \citenamefont {Edlich}, \citenamefont {Davies}, \citenamefont {Newman}, \citenamefont {Jones}, \citenamefont {Satzinger}, \citenamefont {Niu}, \citenamefont {Blackwell}, \citenamefont {Holland}, \citenamefont {Kafri}, \citenamefont {Atalaya}, \citenamefont {Gidney}, \citenamefont {Hassabis}, \citenamefont {Boixo}, \citenamefont {Neven},\ and\ \citenamefont {Kohli}}]{Bausch2023}%
  \BibitemOpen
  \bibfield  {author} {\bibinfo {author} {\bibfnamefont {J.}~\bibnamefont {Bausch}}, \bibinfo {author} {\bibfnamefont {A.~W.}\ \bibnamefont {Senior}}, \bibinfo {author} {\bibfnamefont {F.~J.~H.}\ \bibnamefont {Heras}}, \bibinfo {author} {\bibfnamefont {T.}~\bibnamefont {Edlich}}, \bibinfo {author} {\bibfnamefont {A.}~\bibnamefont {Davies}}, \bibinfo {author} {\bibfnamefont {M.}~\bibnamefont {Newman}}, \bibinfo {author} {\bibfnamefont {C.}~\bibnamefont {Jones}}, \bibinfo {author} {\bibfnamefont {K.}~\bibnamefont {Satzinger}}, \bibinfo {author} {\bibfnamefont {M.~Y.}\ \bibnamefont {Niu}}, \bibinfo {author} {\bibfnamefont {S.}~\bibnamefont {Blackwell}}, \bibinfo {author} {\bibfnamefont {G.}~\bibnamefont {Holland}}, \bibinfo {author} {\bibfnamefont {D.}~\bibnamefont {Kafri}}, \bibinfo {author} {\bibfnamefont {J.}~\bibnamefont {Atalaya}}, \bibinfo {author} {\bibfnamefont {C.}~\bibnamefont {Gidney}}, \bibinfo {author} {\bibfnamefont {D.}~\bibnamefont {Hassabis}}, \bibinfo {author} {\bibfnamefont {S.}~\bibnamefont
  {Boixo}}, \bibinfo {author} {\bibfnamefont {H.}~\bibnamefont {Neven}},\ and\ \bibinfo {author} {\bibfnamefont {P.}~\bibnamefont {Kohli}},\ }\bibfield  {title} {\bibinfo {title} {Learning high-accuracy error decoding for quantum processors},\ }\href {https://doi.org/10.1038/s41586-024-08148-8} {\bibfield  {journal} {\bibinfo  {journal} {Nature}\ }\textbf {\bibinfo {volume} {635}},\ \bibinfo {pages} {834} (\bibinfo {year} {2024})}\BibitemShut {NoStop}%
\bibitem [{\citenamefont {Wen}(2003)}]{Wen2003QuantumOrders}%
  \BibitemOpen
  \bibfield  {author} {\bibinfo {author} {\bibfnamefont {X.-G.}\ \bibnamefont {Wen}},\ }\bibfield  {title} {\bibinfo {title} {{Quantum Orders in an Exact Soluble Model}},\ }\href {https://doi.org/10.1103/PhysRevLett.90.016803} {\bibfield  {journal} {\bibinfo  {journal} {Physical Review Letters}\ }\textbf {\bibinfo {volume} {90}},\ \bibinfo {pages} {016803} (\bibinfo {year} {2003})}\BibitemShut {NoStop}%
\bibitem [{\citenamefont {Bombin}\ and\ \citenamefont {Martin-Delgado}(2007)}]{Bombin2007}%
  \BibitemOpen
  \bibfield  {author} {\bibinfo {author} {\bibfnamefont {H.}~\bibnamefont {Bombin}}\ and\ \bibinfo {author} {\bibfnamefont {M.~A.}\ \bibnamefont {Martin-Delgado}},\ }\bibfield  {title} {\bibinfo {title} {{Optimal resources for topological two-dimensional stabilizer codes: Comparative study}},\ }\href {https://doi.org/10.1103/physreva.76.012305} {\bibfield  {journal} {\bibinfo  {journal} {Physical Review A}\ }\textbf {\bibinfo {volume} {76}},\ \bibinfo {pages} {012305} (\bibinfo {year} {2007})}\BibitemShut {NoStop}%
\bibitem [{\citenamefont {Kay}(2011)}]{Kay2011Capabilities}%
  \BibitemOpen
  \bibfield  {author} {\bibinfo {author} {\bibfnamefont {A.}~\bibnamefont {Kay}},\ }\bibfield  {title} {\bibinfo {title} {{Capabilities of a Perturbed Toric Code as a Quantum Memory}},\ }\href {https://doi.org/10.1103/PhysRevLett.107.270502} {\bibfield  {journal} {\bibinfo  {journal} {Physical Review Letters}\ }\textbf {\bibinfo {volume} {107}},\ \bibinfo {pages} {270502} (\bibinfo {year} {2011})}\BibitemShut {NoStop}%
\bibitem [{\citenamefont {Shulman}\ \emph {et~al.}(2012)\citenamefont {Shulman}, \citenamefont {Dial}, \citenamefont {Harvey}, \citenamefont {Bluhm}, \citenamefont {Umansky},\ and\ \citenamefont {Yacoby}}]{Shulman2012}%
  \BibitemOpen
  \bibfield  {author} {\bibinfo {author} {\bibfnamefont {M.~D.}\ \bibnamefont {Shulman}}, \bibinfo {author} {\bibfnamefont {O.~E.}\ \bibnamefont {Dial}}, \bibinfo {author} {\bibfnamefont {S.~P.}\ \bibnamefont {Harvey}}, \bibinfo {author} {\bibfnamefont {H.}~\bibnamefont {Bluhm}}, \bibinfo {author} {\bibfnamefont {V.}~\bibnamefont {Umansky}},\ and\ \bibinfo {author} {\bibfnamefont {A.}~\bibnamefont {Yacoby}},\ }\bibfield  {title} {\bibinfo {title} {{Demonstration of Entanglement of Electrostatically Coupled Singlet-Triplet Qubits}},\ }\href {https://doi.org/10.1126/science.1217692} {\bibfield  {journal} {\bibinfo  {journal} {Science}\ }\textbf {\bibinfo {volume} {336}},\ \bibinfo {pages} {202} (\bibinfo {year} {2012})}\BibitemShut {NoStop}%
\bibitem [{\citenamefont {Pop}\ \emph {et~al.}(2014)\citenamefont {Pop}, \citenamefont {Geerlings}, \citenamefont {Catelani}, \citenamefont {Schoelkopf}, \citenamefont {Glazman},\ and\ \citenamefont {Devoret}}]{Pop2014}%
  \BibitemOpen
  \bibfield  {author} {\bibinfo {author} {\bibfnamefont {I.~M.}\ \bibnamefont {Pop}}, \bibinfo {author} {\bibfnamefont {K.}~\bibnamefont {Geerlings}}, \bibinfo {author} {\bibfnamefont {G.}~\bibnamefont {Catelani}}, \bibinfo {author} {\bibfnamefont {R.~J.}\ \bibnamefont {Schoelkopf}}, \bibinfo {author} {\bibfnamefont {L.~I.}\ \bibnamefont {Glazman}},\ and\ \bibinfo {author} {\bibfnamefont {M.~H.}\ \bibnamefont {Devoret}},\ }\bibfield  {title} {\bibinfo {title} {Coherent suppression of electromagnetic dissipation due to superconducting quasiparticles},\ }\href {https://doi.org/10.1038/nature13017} {\bibfield  {journal} {\bibinfo  {journal} {Nature}\ }\textbf {\bibinfo {volume} {508}},\ \bibinfo {pages} {369} (\bibinfo {year} {2014})}\BibitemShut {NoStop}%
\bibitem [{\citenamefont {Waldherr}\ \emph {et~al.}(2014)\citenamefont {Waldherr}, \citenamefont {Wang}, \citenamefont {Zaiser}, \citenamefont {Jamali}, \citenamefont {Schulte-Herbr\"{u}ggen}, \citenamefont {Abe}, \citenamefont {Ohshima}, \citenamefont {Isoya}, \citenamefont {Du}, \citenamefont {Neumann},\ and\ \citenamefont {Wrachtrup}}]{Waldherr2014}%
  \BibitemOpen
  \bibfield  {author} {\bibinfo {author} {\bibfnamefont {G.}~\bibnamefont {Waldherr}}, \bibinfo {author} {\bibfnamefont {Y.}~\bibnamefont {Wang}}, \bibinfo {author} {\bibfnamefont {S.}~\bibnamefont {Zaiser}}, \bibinfo {author} {\bibfnamefont {M.}~\bibnamefont {Jamali}}, \bibinfo {author} {\bibfnamefont {T.}~\bibnamefont {Schulte-Herbr\"{u}ggen}}, \bibinfo {author} {\bibfnamefont {H.}~\bibnamefont {Abe}}, \bibinfo {author} {\bibfnamefont {T.}~\bibnamefont {Ohshima}}, \bibinfo {author} {\bibfnamefont {J.}~\bibnamefont {Isoya}}, \bibinfo {author} {\bibfnamefont {J.~F.}\ \bibnamefont {Du}}, \bibinfo {author} {\bibfnamefont {P.}~\bibnamefont {Neumann}},\ and\ \bibinfo {author} {\bibfnamefont {J.}~\bibnamefont {Wrachtrup}},\ }\bibfield  {title} {\bibinfo {title} {Quantum error correction in a solid-state hybrid spin register},\ }\href {https://doi.org/10.1038/nature12919} {\bibfield  {journal} {\bibinfo  {journal} {Nature}\ }\textbf {\bibinfo {volume} {506}},\ \bibinfo {pages} {204} (\bibinfo {year}
  {2014})}\BibitemShut {NoStop}%
\bibitem [{\citenamefont {Watson}\ \emph {et~al.}(2018)\citenamefont {Watson}, \citenamefont {Philips}, \citenamefont {Kawakami}, \citenamefont {Ward}, \citenamefont {Scarlino}, \citenamefont {Veldhorst}, \citenamefont {Savage}, \citenamefont {Lagally}, \citenamefont {Friesen}, \citenamefont {Coppersmith}, \citenamefont {Eriksson},\ and\ \citenamefont {Vandersypen}}]{Watson2018}%
  \BibitemOpen
  \bibfield  {author} {\bibinfo {author} {\bibfnamefont {T.~F.}\ \bibnamefont {Watson}}, \bibinfo {author} {\bibfnamefont {S.~G.~J.}\ \bibnamefont {Philips}}, \bibinfo {author} {\bibfnamefont {E.}~\bibnamefont {Kawakami}}, \bibinfo {author} {\bibfnamefont {D.~R.}\ \bibnamefont {Ward}}, \bibinfo {author} {\bibfnamefont {P.}~\bibnamefont {Scarlino}}, \bibinfo {author} {\bibfnamefont {M.}~\bibnamefont {Veldhorst}}, \bibinfo {author} {\bibfnamefont {D.~E.}\ \bibnamefont {Savage}}, \bibinfo {author} {\bibfnamefont {M.~G.}\ \bibnamefont {Lagally}}, \bibinfo {author} {\bibfnamefont {M.}~\bibnamefont {Friesen}}, \bibinfo {author} {\bibfnamefont {S.~N.}\ \bibnamefont {Coppersmith}}, \bibinfo {author} {\bibfnamefont {M.~A.}\ \bibnamefont {Eriksson}},\ and\ \bibinfo {author} {\bibfnamefont {L.~M.~K.}\ \bibnamefont {Vandersypen}},\ }\bibfield  {title} {\bibinfo {title} {A programmable two-qubit quantum processor in silicon},\ }\href {https://doi.org/10.1038/nature25766} {\bibfield  {journal} {\bibinfo  {journal}
  {Nature}\ }\textbf {\bibinfo {volume} {555}},\ \bibinfo {pages} {633} (\bibinfo {year} {2018})}\BibitemShut {NoStop}%
\bibitem [{\citenamefont {Lescanne}\ \emph {et~al.}(2020)\citenamefont {Lescanne}, \citenamefont {Villiers}, \citenamefont {Peronnin}, \citenamefont {Sarlette}, \citenamefont {Delbecq}, \citenamefont {Huard}, \citenamefont {Kontos}, \citenamefont {Mirrahimi},\ and\ \citenamefont {Leghtas}}]{Lescanne2020}%
  \BibitemOpen
  \bibfield  {author} {\bibinfo {author} {\bibfnamefont {R.}~\bibnamefont {Lescanne}}, \bibinfo {author} {\bibfnamefont {M.}~\bibnamefont {Villiers}}, \bibinfo {author} {\bibfnamefont {T.}~\bibnamefont {Peronnin}}, \bibinfo {author} {\bibfnamefont {A.}~\bibnamefont {Sarlette}}, \bibinfo {author} {\bibfnamefont {M.}~\bibnamefont {Delbecq}}, \bibinfo {author} {\bibfnamefont {B.}~\bibnamefont {Huard}}, \bibinfo {author} {\bibfnamefont {T.}~\bibnamefont {Kontos}}, \bibinfo {author} {\bibfnamefont {M.}~\bibnamefont {Mirrahimi}},\ and\ \bibinfo {author} {\bibfnamefont {Z.}~\bibnamefont {Leghtas}},\ }\bibfield  {title} {\bibinfo {title} {Exponential suppression of bit-flips in a qubit encoded in an oscillator},\ }\href {https://doi.org/10.1038/s41567-020-0824-x} {\bibfield  {journal} {\bibinfo  {journal} {Nature Physics}\ }\textbf {\bibinfo {volume} {16}},\ \bibinfo {pages} {509} (\bibinfo {year} {2020})}\BibitemShut {NoStop}%
\bibitem [{\citenamefont {H\"{a}nggli}\ \emph {et~al.}(2020)\citenamefont {H\"{a}nggli}, \citenamefont {Heinze},\ and\ \citenamefont {K\"{o}nig}}]{Hnggli2020}%
  \BibitemOpen
  \bibfield  {author} {\bibinfo {author} {\bibfnamefont {L.}~\bibnamefont {H\"{a}nggli}}, \bibinfo {author} {\bibfnamefont {M.}~\bibnamefont {Heinze}},\ and\ \bibinfo {author} {\bibfnamefont {R.}~\bibnamefont {K\"{o}nig}},\ }\bibfield  {title} {\bibinfo {title} {{Enhanced noise resilience of the surface–Gottesman-Kitaev-Preskill code via designed bias}},\ }\href {https://doi.org/10.1103/physreva.102.052408} {\bibfield  {journal} {\bibinfo  {journal} {Physical Review A}\ }\textbf {\bibinfo {volume} {102}},\ \bibinfo {pages} {052408} (\bibinfo {year} {2020})}\BibitemShut {NoStop}%
\bibitem [{\citenamefont {Hajr}\ \emph {et~al.}(2024)\citenamefont {Hajr}, \citenamefont {Qing}, \citenamefont {Wang}, \citenamefont {Koolstra}, \citenamefont {Pedramrazi}, \citenamefont {Kang}, \citenamefont {Chen}, \citenamefont {Nguyen}, \citenamefont {Junger}, \citenamefont {Goss}, \citenamefont {Huang}, \citenamefont {Bhandari}, \citenamefont {Frattini}, \citenamefont {Puri}, \citenamefont {Dressel}, \citenamefont {Jordan}, \citenamefont {Santiago},\ and\ \citenamefont {Siddiqi}}]{Hajr2024}%
  \BibitemOpen
  \bibfield  {author} {\bibinfo {author} {\bibfnamefont {A.}~\bibnamefont {Hajr}}, \bibinfo {author} {\bibfnamefont {B.}~\bibnamefont {Qing}}, \bibinfo {author} {\bibfnamefont {K.}~\bibnamefont {Wang}}, \bibinfo {author} {\bibfnamefont {G.}~\bibnamefont {Koolstra}}, \bibinfo {author} {\bibfnamefont {Z.}~\bibnamefont {Pedramrazi}}, \bibinfo {author} {\bibfnamefont {Z.}~\bibnamefont {Kang}}, \bibinfo {author} {\bibfnamefont {L.}~\bibnamefont {Chen}}, \bibinfo {author} {\bibfnamefont {L.~B.}\ \bibnamefont {Nguyen}}, \bibinfo {author} {\bibfnamefont {C.}~\bibnamefont {Junger}}, \bibinfo {author} {\bibfnamefont {N.}~\bibnamefont {Goss}}, \bibinfo {author} {\bibfnamefont {I.}~\bibnamefont {Huang}}, \bibinfo {author} {\bibfnamefont {B.}~\bibnamefont {Bhandari}}, \bibinfo {author} {\bibfnamefont {N.~E.}\ \bibnamefont {Frattini}}, \bibinfo {author} {\bibfnamefont {S.}~\bibnamefont {Puri}}, \bibinfo {author} {\bibfnamefont {J.}~\bibnamefont {Dressel}}, \bibinfo {author} {\bibfnamefont {A.~N.}\ \bibnamefont {Jordan}},
  \bibinfo {author} {\bibfnamefont {D.}~\bibnamefont {Santiago}},\ and\ \bibinfo {author} {\bibfnamefont {I.}~\bibnamefont {Siddiqi}},\ }\bibfield  {title} {\bibinfo {title} {{High-Coherence Kerr-cat qubit in 2D architecture}},\ }\href {https://doi.org/10.1103/PhysRevX.14.041049} {\bibfield  {journal} {\bibinfo  {journal} {Physical Review X}\ }\textbf {\bibinfo {volume} {14}},\ \bibinfo {pages} {041049} (\bibinfo {year} {2024})}\BibitemShut {NoStop}%
\bibitem [{\citenamefont {Cong}\ \emph {et~al.}(2022)\citenamefont {Cong}, \citenamefont {Levine}, \citenamefont {Keesling}, \citenamefont {Bluvstein}, \citenamefont {Wang},\ and\ \citenamefont {Lukin}}]{Cong2022}%
  \BibitemOpen
  \bibfield  {author} {\bibinfo {author} {\bibfnamefont {I.}~\bibnamefont {Cong}}, \bibinfo {author} {\bibfnamefont {H.}~\bibnamefont {Levine}}, \bibinfo {author} {\bibfnamefont {A.}~\bibnamefont {Keesling}}, \bibinfo {author} {\bibfnamefont {D.}~\bibnamefont {Bluvstein}}, \bibinfo {author} {\bibfnamefont {S.-T.}\ \bibnamefont {Wang}},\ and\ \bibinfo {author} {\bibfnamefont {M.~D.}\ \bibnamefont {Lukin}},\ }\bibfield  {title} {\bibinfo {title} {{Hardware-Efficient, Fault-Tolerant Quantum Computation with Rydberg Atoms}},\ }\href {https://doi.org/10.1103/PhysRevX.12.021049} {\bibfield  {journal} {\bibinfo  {journal} {Physical Review X}\ }\textbf {\bibinfo {volume} {12}},\ \bibinfo {pages} {021049} (\bibinfo {year} {2022})}\BibitemShut {NoStop}%
\bibitem [{\citenamefont {Puri}\ \emph {et~al.}(2020)\citenamefont {Puri}, \citenamefont {St-Jean}, \citenamefont {Gross}, \citenamefont {Grimm}, \citenamefont {Frattini}, \citenamefont {Iyer}, \citenamefont {Krishna}, \citenamefont {Touzard}, \citenamefont {Jiang}, \citenamefont {Blais}, \citenamefont {Flammia},\ and\ \citenamefont {Girvin}}]{Puri2020}%
  \BibitemOpen
  \bibfield  {author} {\bibinfo {author} {\bibfnamefont {S.}~\bibnamefont {Puri}}, \bibinfo {author} {\bibfnamefont {L.}~\bibnamefont {St-Jean}}, \bibinfo {author} {\bibfnamefont {J.~A.}\ \bibnamefont {Gross}}, \bibinfo {author} {\bibfnamefont {A.}~\bibnamefont {Grimm}}, \bibinfo {author} {\bibfnamefont {N.~E.}\ \bibnamefont {Frattini}}, \bibinfo {author} {\bibfnamefont {P.~S.}\ \bibnamefont {Iyer}}, \bibinfo {author} {\bibfnamefont {A.}~\bibnamefont {Krishna}}, \bibinfo {author} {\bibfnamefont {S.}~\bibnamefont {Touzard}}, \bibinfo {author} {\bibfnamefont {L.}~\bibnamefont {Jiang}}, \bibinfo {author} {\bibfnamefont {A.}~\bibnamefont {Blais}}, \bibinfo {author} {\bibfnamefont {S.~T.}\ \bibnamefont {Flammia}},\ and\ \bibinfo {author} {\bibfnamefont {S.~M.}\ \bibnamefont {Girvin}},\ }\bibfield  {title} {\bibinfo {title} {Bias-preserving gates with stabilized cat qubits},\ }\href {https://doi.org/10.1126/sciadv.aay5901} {\bibfield  {journal} {\bibinfo  {journal} {Science Advances}\ }\textbf {\bibinfo {volume}
  {6}},\ \bibinfo {pages} {eaay5901} (\bibinfo {year} {2020})}\BibitemShut {NoStop}%
\bibitem [{\citenamefont {Darmawan}\ \emph {et~al.}(2021)\citenamefont {Darmawan}, \citenamefont {Brown}, \citenamefont {Grimsmo}, \citenamefont {Tuckett},\ and\ \citenamefont {Puri}}]{Darmawan2021}%
  \BibitemOpen
  \bibfield  {author} {\bibinfo {author} {\bibfnamefont {A.~S.}\ \bibnamefont {Darmawan}}, \bibinfo {author} {\bibfnamefont {B.~J.}\ \bibnamefont {Brown}}, \bibinfo {author} {\bibfnamefont {A.~L.}\ \bibnamefont {Grimsmo}}, \bibinfo {author} {\bibfnamefont {D.~K.}\ \bibnamefont {Tuckett}},\ and\ \bibinfo {author} {\bibfnamefont {S.}~\bibnamefont {Puri}},\ }\bibfield  {title} {\bibinfo {title} {{Practical Quantum Error Correction with the XZZX Code and Kerr-Cat Qubits}},\ }\href {https://doi.org/10.1103/prxquantum.2.030345} {\bibfield  {journal} {\bibinfo  {journal} {PRX Quantum}\ }\textbf {\bibinfo {volume} {2}},\ \bibinfo {pages} {030345} (\bibinfo {year} {2021})}\BibitemShut {NoStop}%
\bibitem [{\citenamefont {Miguel}\ \emph {et~al.}(2023)\citenamefont {Miguel}, \citenamefont {Williamson},\ and\ \citenamefont {Brown}}]{Miguel2023}%
  \BibitemOpen
  \bibfield  {author} {\bibinfo {author} {\bibfnamefont {J.~F.~S.}\ \bibnamefont {Miguel}}, \bibinfo {author} {\bibfnamefont {D.~J.}\ \bibnamefont {Williamson}},\ and\ \bibinfo {author} {\bibfnamefont {B.~J.}\ \bibnamefont {Brown}},\ }\bibfield  {title} {\bibinfo {title} {A cellular automaton decoder for a noise-bias tailored color code},\ }\href {https://doi.org/10.22331/q-2023-03-09-940} {\bibfield  {journal} {\bibinfo  {journal} {Quantum}\ }\textbf {\bibinfo {volume} {7}},\ \bibinfo {pages} {940} (\bibinfo {year} {2023})}\BibitemShut {NoStop}%
\bibitem [{\citenamefont {Guillaud}\ and\ \citenamefont {Mirrahimi}(2019)}]{Guillaud2019}%
  \BibitemOpen
  \bibfield  {author} {\bibinfo {author} {\bibfnamefont {J.}~\bibnamefont {Guillaud}}\ and\ \bibinfo {author} {\bibfnamefont {M.}~\bibnamefont {Mirrahimi}},\ }\bibfield  {title} {\bibinfo {title} {{Repetition Cat Qubits for Fault-Tolerant Quantum Computation}},\ }\href {https://doi.org/10.1103/physrevx.9.041053} {\bibfield  {journal} {\bibinfo  {journal} {Physical Review X}\ }\textbf {\bibinfo {volume} {9}},\ \bibinfo {pages} {041053} (\bibinfo {year} {2019})}\BibitemShut {NoStop}%
\bibitem [{\citenamefont {Claes}\ \emph {et~al.}(2023)\citenamefont {Claes}, \citenamefont {Bourassa},\ and\ \citenamefont {Puri}}]{Claes2023}%
  \BibitemOpen
  \bibfield  {author} {\bibinfo {author} {\bibfnamefont {J.}~\bibnamefont {Claes}}, \bibinfo {author} {\bibfnamefont {J.~E.}\ \bibnamefont {Bourassa}},\ and\ \bibinfo {author} {\bibfnamefont {S.}~\bibnamefont {Puri}},\ }\bibfield  {title} {\bibinfo {title} {Tailored cluster states with high threshold under biased noise},\ }\href {https://doi.org/10.1038/s41534-023-00677-w} {\bibfield  {journal} {\bibinfo  {journal} {npj Quantum Information}\ }\textbf {\bibinfo {volume} {9}},\ \bibinfo {pages} {9} (\bibinfo {year} {2023})}\BibitemShut {NoStop}%
\bibitem [{\citenamefont {Piveteau}\ \emph {et~al.}(2024)\citenamefont {Piveteau}, \citenamefont {Chubb},\ and\ \citenamefont {Renes}}]{PRXQuantum.5.040303}%
  \BibitemOpen
  \bibfield  {author} {\bibinfo {author} {\bibfnamefont {C.}~\bibnamefont {Piveteau}}, \bibinfo {author} {\bibfnamefont {C.~T.}\ \bibnamefont {Chubb}},\ and\ \bibinfo {author} {\bibfnamefont {J.~M.}\ \bibnamefont {Renes}},\ }\bibfield  {title} {\bibinfo {title} {{Tensor-Network Decoding Beyond 2D}},\ }\href {https://doi.org/10.1103/PRXQuantum.5.040303} {\bibfield  {journal} {\bibinfo  {journal} {PRX Quantum}\ }\textbf {\bibinfo {volume} {5}},\ \bibinfo {pages} {040303} (\bibinfo {year} {2024})}\BibitemShut {NoStop}%
\bibitem [{\citenamefont {Chou}\ \emph {et~al.}(2024)\citenamefont {Chou} \emph {et~al.}}]{Chou2024}%
  \BibitemOpen
  \bibfield  {author} {\bibinfo {author} {\bibfnamefont {K.~S.}\ \bibnamefont {Chou}} \emph {et~al.},\ }\bibfield  {title} {\bibinfo {title} {A superconducting dual-rail cavity qubit with erasure-detected logical measurements},\ }\href {https://doi.org/10.1038/s41567-024-02539-4} {\bibfield  {journal} {\bibinfo  {journal} {Nature Physics}\ }\textbf {\bibinfo {volume} {20}},\ \bibinfo {pages} {1454} (\bibinfo {year} {2024})}\BibitemShut {NoStop}%
\bibitem [{\citenamefont {Teoh}\ \emph {et~al.}(2023)\citenamefont {Teoh}, \citenamefont {Winkel}, \citenamefont {Babla}, \citenamefont {Chapman}, \citenamefont {Claes}, \citenamefont {de~Graaf}, \citenamefont {Garmon}, \citenamefont {Kalfus}, \citenamefont {Lu}, \citenamefont {Maiti}, \citenamefont {Sahay}, \citenamefont {Thakur}, \citenamefont {Tsunoda}, \citenamefont {Xue}, \citenamefont {Frunzio}, \citenamefont {Girvin}, \citenamefont {Puri},\ and\ \citenamefont {Schoelkopf}}]{Teoh2023}%
  \BibitemOpen
  \bibfield  {author} {\bibinfo {author} {\bibfnamefont {J.~D.}\ \bibnamefont {Teoh}}, \bibinfo {author} {\bibfnamefont {P.}~\bibnamefont {Winkel}}, \bibinfo {author} {\bibfnamefont {H.~K.}\ \bibnamefont {Babla}}, \bibinfo {author} {\bibfnamefont {B.~J.}\ \bibnamefont {Chapman}}, \bibinfo {author} {\bibfnamefont {J.}~\bibnamefont {Claes}}, \bibinfo {author} {\bibfnamefont {S.~J.}\ \bibnamefont {de~Graaf}}, \bibinfo {author} {\bibfnamefont {J.~W.~O.}\ \bibnamefont {Garmon}}, \bibinfo {author} {\bibfnamefont {W.~D.}\ \bibnamefont {Kalfus}}, \bibinfo {author} {\bibfnamefont {Y.}~\bibnamefont {Lu}}, \bibinfo {author} {\bibfnamefont {A.}~\bibnamefont {Maiti}}, \bibinfo {author} {\bibfnamefont {K.}~\bibnamefont {Sahay}}, \bibinfo {author} {\bibfnamefont {N.}~\bibnamefont {Thakur}}, \bibinfo {author} {\bibfnamefont {T.}~\bibnamefont {Tsunoda}}, \bibinfo {author} {\bibfnamefont {S.~H.}\ \bibnamefont {Xue}}, \bibinfo {author} {\bibfnamefont {L.}~\bibnamefont {Frunzio}}, \bibinfo {author} {\bibfnamefont {S.~M.}\
  \bibnamefont {Girvin}}, \bibinfo {author} {\bibfnamefont {S.}~\bibnamefont {Puri}},\ and\ \bibinfo {author} {\bibfnamefont {R.~J.}\ \bibnamefont {Schoelkopf}},\ }\bibfield  {title} {\bibinfo {title} {Dual-rail encoding with superconducting cavities},\ }\href {https://doi.org/10.1073/pnas.2221736120} {\bibfield  {journal} {\bibinfo  {journal} {Proceedings of the National Academy of Sciences}\ }\textbf {\bibinfo {volume} {120}},\ \bibinfo {pages} {e2221736120} (\bibinfo {year} {2023})}\BibitemShut {NoStop}%
\bibitem [{\citenamefont {Koottandavida}\ \emph {et~al.}(2024)\citenamefont {Koottandavida}, \citenamefont {Tsioutsios}, \citenamefont {Kargioti}, \citenamefont {Smith}, \citenamefont {Joshi}, \citenamefont {Dai}, \citenamefont {Teoh}, \citenamefont {Curtis}, \citenamefont {Frunzio}, \citenamefont {Schoelkopf},\ and\ \citenamefont {Devoret}}]{Koottandavida2024}%
  \BibitemOpen
  \bibfield  {author} {\bibinfo {author} {\bibfnamefont {A.}~\bibnamefont {Koottandavida}}, \bibinfo {author} {\bibfnamefont {I.}~\bibnamefont {Tsioutsios}}, \bibinfo {author} {\bibfnamefont {A.}~\bibnamefont {Kargioti}}, \bibinfo {author} {\bibfnamefont {C.~R.}\ \bibnamefont {Smith}}, \bibinfo {author} {\bibfnamefont {V.~R.}\ \bibnamefont {Joshi}}, \bibinfo {author} {\bibfnamefont {W.}~\bibnamefont {Dai}}, \bibinfo {author} {\bibfnamefont {J.~D.}\ \bibnamefont {Teoh}}, \bibinfo {author} {\bibfnamefont {J.~C.}\ \bibnamefont {Curtis}}, \bibinfo {author} {\bibfnamefont {L.}~\bibnamefont {Frunzio}}, \bibinfo {author} {\bibfnamefont {R.~J.}\ \bibnamefont {Schoelkopf}},\ and\ \bibinfo {author} {\bibfnamefont {M.~H.}\ \bibnamefont {Devoret}},\ }\bibfield  {title} {\bibinfo {title} {{Erasure Detection of a Dual-Rail Qubit Encoded in a Double-Post Superconducting Cavity}},\ }\href {https://doi.org/10.1103/PhysRevLett.132.180601} {\bibfield  {journal} {\bibinfo  {journal} {Physical Review Letters}\ }\textbf {\bibinfo
  {volume} {132}},\ \bibinfo {pages} {180601} (\bibinfo {year} {2024})}\BibitemShut {NoStop}%
\bibitem [{\citenamefont {Levine}\ \emph {et~al.}(2024)\citenamefont {Levine} \emph {et~al.}}]{Levine2024}%
  \BibitemOpen
  \bibfield  {author} {\bibinfo {author} {\bibfnamefont {H.}~\bibnamefont {Levine}} \emph {et~al.},\ }\bibfield  {title} {\bibinfo {title} {{Demonstrating a Long-Coherence Dual-Rail Erasure Qubit Using Tunable Transmons}},\ }\href {https://doi.org/10.1103/physrevx.14.011051} {\bibfield  {journal} {\bibinfo  {journal} {Physical Review X}\ }\textbf {\bibinfo {volume} {14}},\ \bibinfo {pages} {011051} (\bibinfo {year} {2024})}\BibitemShut {NoStop}%
\end{thebibliography}%

\appendix

\section{Comparing our work to a non-sequential matching decoder}\label{app:bence-decoder}

\begin{figure}\centering
\includegraphics[width=\linewidth, trim={0.3cm 0.2cm 0.9cm 0.85cm},clip]{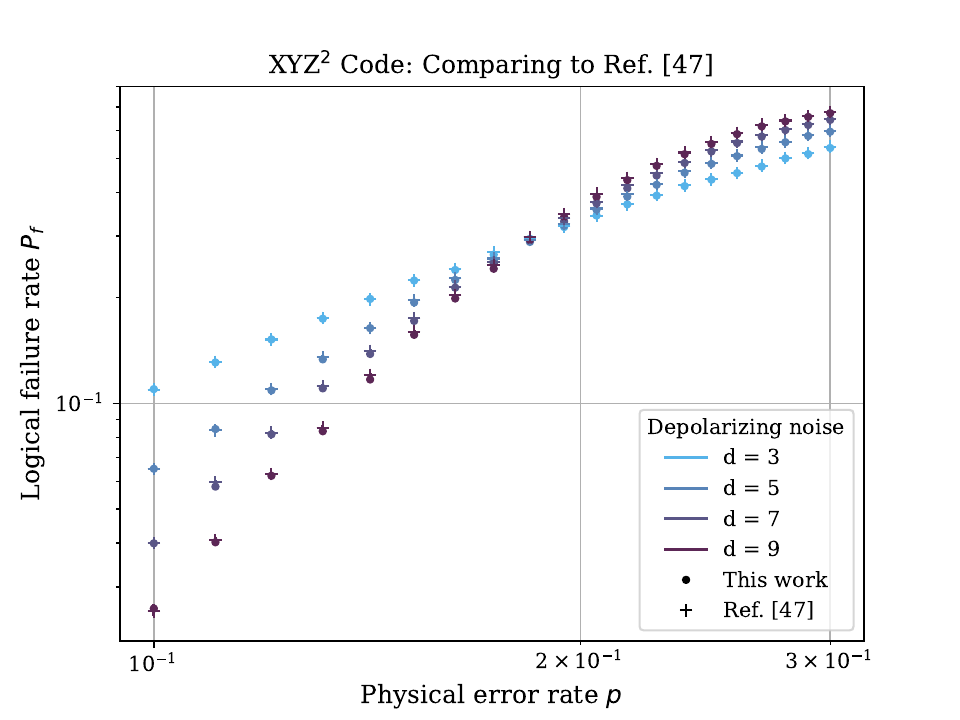}
\caption{Logical failure rate $P_f$ as a function of the physical error rate $p$ for different distances of the \xyz code, under a code-capacity depolarizing noise error model ($\eta = 0.5$). Each data point is evaluated using $5 \times 10^4$ syndromes, decoded using the sequential decoding scheme with a matching decoder on the higher-level code (circular marker), and the decoding scheme with a matching decoder used in Ref.~\cite{hetenyi2024tailoring} (plus marker). We observe thresholds of $18.51 \pm \SI{0.08}{\percent}$ for the matching decoder of our work, compared to $18.27 \pm \SI{0.08}{\percent}$ for the decoder in Ref.~\cite{hetenyi2024tailoring}.}
\label{fig:compare-dep}
\end{figure}

\begin{figure}\centering
\includegraphics[width=\linewidth, trim={0.3cm 0.2cm 0.9cm 0.85cm},clip]{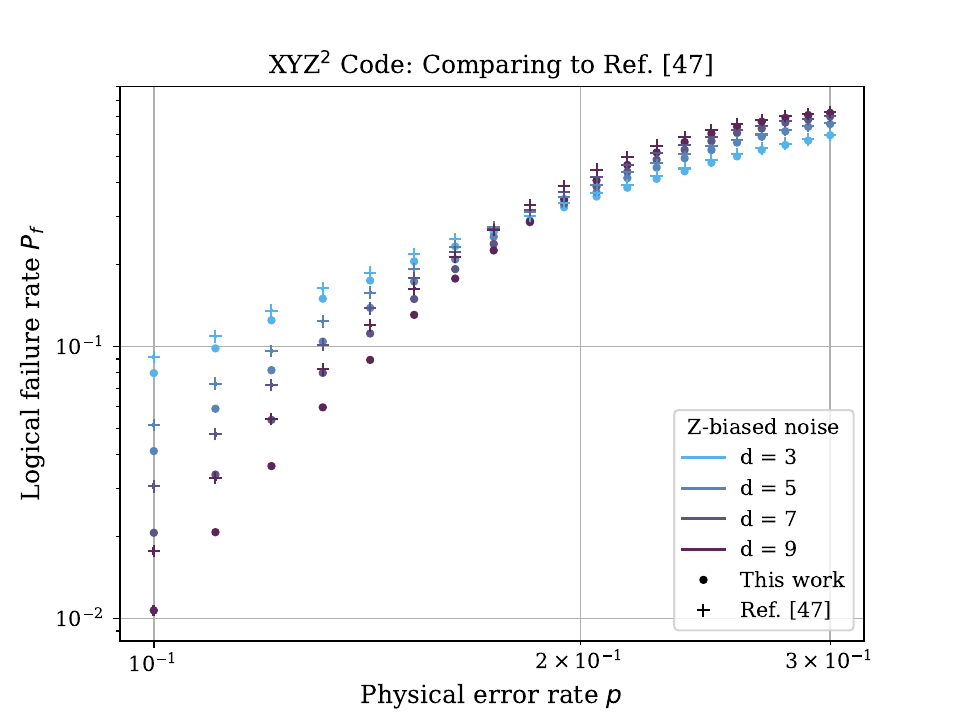}
\caption{Logical failure rate $P_f$ as a function of the physical error rate $p$ for different distances of the \xyz code, under a code-capacity $Z$-biased noise error model ($\eta = 10$). Each data point is evaluated using $5 \times 10^4$ syndromes, decoded using the sequential decoding scheme with a matching decoder on the higher-level code (circular marker), and the decoding scheme with a matching decoder used in Ref.~\cite{hetenyi2024tailoring} (plus marker). We observe thresholds of $18.64 \pm \SI{0.09}{\percent}$ for the matching decoder of our work, compared to $17.46 \pm \SI{0.09}{\percent}$ for the decoder in Ref.~\cite{hetenyi2024tailoring}.}
\label{fig:compare-bias}
\end{figure}

We compare the performance of our sequential decoding scheme using a matching decoder on the higher-level code to decode the \xyz code, with the decoder presented in Ref.~\cite{hetenyi2024tailoring}. The latter matches directly on the \xyz code by duplicating single link syndromes into pairs that match to separate plaquette sublattices.  In \figref{fig:compare-dep}, we present the comparison between the two decoders for code-capacity depolarizing noise, decoding the same set of simulated errors with both decoders. We also study the performance under code-capacity biased noise (with $\eta = 10$) in \figref{fig:compare-bias}. We observe a slight increase in the performance of our decoder for biased noise. We attribute this to the inherent modification of the error model when decoding using the sequential decoding scheme, as mentioned in \secref{subsec:bias-mapping}. We note that we only study the two decoders under code-capacity noise for a fair comparison, as the inherent structure of the circuit in Ref.~\cite{hetenyi2024tailoring} is different from the phenomenological noise model used in this work.

\section{Performance of belief-matching and matching under phenomenological noise}\label{app:bm-vs-matching}

We studied the performance of the sequential decoding scheme for the \xyz code for phenomenological noise using belief matching and matching as the decoder for the higher-level YZZY code. As seen in \figref{fig:phen-noise-matching-dep}, we find that belief matching has slightly higher logical failure rates compared to matching for the distances and physical error rates evaluated, even though the thresholds are roughly within error bars. This behaviour can be attributed to a few rare high-weight error configurations, which together with our choice of decoding on the lower- and higher-level codes lead to a logical failure when using belief matching on the higher-level code, but not when using matching.

To decode the \xyz code under phenomenological noise, we first decode the link syndromes. This is a simplified decoding strategy that does not take into account any extra information that would be available from the neighbouring plaquette stabilizers. In addition, we correct the link syndromes by placing a Pauli $Z$ correction on the top qubit of the links with non-trivial syndromes in the cases where they are not interpreted as measurement errors.

In some cases, one of the pair of non-trivial detectors caused by measurement errors gets paired with a link syndrome caused by a data-qubit error. In this case, the other pair of the link detector due to measurement error survives and leads to the placement of $Z$ corrections incorrectly on link syndromes that were not due to data errors.
As the placement of these Paulis is followed by modifying the corresponding plaquette stabilizers and the priors on the respective higher-level code qubits accordingly, the information being fed into the belief-propagation subroutine is faulty. As a result, belief propagation outputs higher priors on qubits other than the original error location. This, in turn, leads the matching decoder to match the syndromes incorrectly, in some cases leading to a logical failure. In contrast, using a bare matching decoder either gets uniform priors on the code qubits based on the underlying physical error rate, or an indication of where the erasure errors might have occurred as signified by the placement of the Pauli $Z$ correction.

\begin{figure*}\centering
\includegraphics[width=0.85\linewidth, trim={0cm 0cm 0cm 0cm},clip]{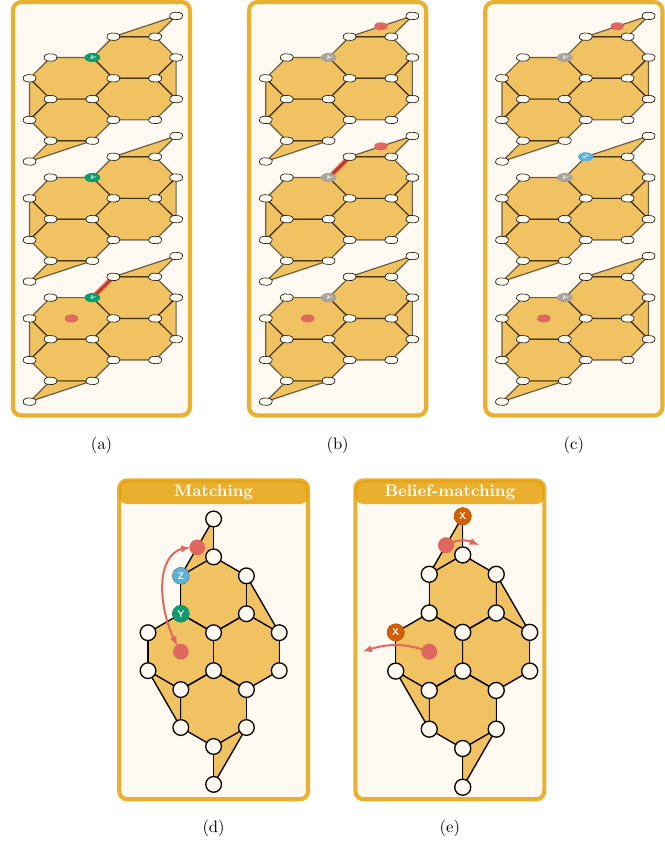}
\caption{Example of a weight-3 error on a distance-3 \xyz code that leads to a logical failure when using belief matching but decodes correctly using a matching decoder. (a) Initial Pauli error in the first time step that gives rise to detectors as shown. (b) Measurement errors that occur while measuring the syndromes: one on a link that \textit{moves} the original link detector and a second one that causes two plaquette detectors on the boundary to light up. (c) First decoding step that places a $Z$ correction to get rid of the link syndrome. The next steps will decode the remaining plaquette syndromes. (d) Top view showing the direction in which the matching decoder matches the remaining plaquette syndromes. This is equivalent up to the error that originally occurred and the correction we placed in the previous step and so the logical equivalence class is preserved. (e) Top view showing the direction in which the belief-matching decoder matches the remaining plaquette syndromes. This happens because the belief-propagation subroutine increases the beliefs on the qubits next to the violated plaquette detectors. The correction along with the original error causes a logical failure in this case.}
\label{fig:bp-vs-m}
\end{figure*}

A particular example of such an error configuration is shown in \figref{fig:bp-vs-m}, where the syndrome fails to be decoded correctly using the belief-matching decoder, but is decoded correctly using a matching decoder.

\section{Proof of optimal sequential decoding}\label{app:proof-optimal-decoding}

Although the general proof that the sequential decoding scheme is optimal, given a maximum-likelihood decoder for the high-level YZZY code, is given in Ref.~\cite{Poulin2006Optimal}, we present an informal derivation for our decoding algorithm. To construct an optimal decoder, we need to identify the most likely equivalence class of errors that corresponds to any given syndrome, which is achieved by the following steps:  
\begin{enumerate}
    \item On each link, identify error chains on the \xyz code that differ only by the action of the corresponding link stabilizer. Acting with a stabilizer will not change the equivalence class of the chain. Thus, for example, $II$ and $XX$ are identified as a single error event, with probability $p_i^2+p_x^2$.
    \item On each link, there are two sets of four possible error configurations, as specified in \tabref{table:prior-update-rule}, distinguished by whether there is a link syndrome or not and by the corresponding Pauli error (including no error) on the higher-level code.
    \item Next, map any syndrome with one or more triggered link stabilizers to a corresponding syndrome without triggered links, in a way that conserves the probability of all error chains and does not change the equivalence classes. To do this, we ``correct'' each triggered link by acting with $Z$ on the top qubit. (The choice of which qubit and if we use $Z$ or $Y$ is arbitrary, but \tabref{table:prior-update-rule} reflects this particular choice.)
    \item The set of $Z$ operators on the corrected links either commute or anticommute with the logical operators (depending on the choice of the representation of the logical operator). Since all the error chains consistent with the syndrome are multiplied with the same set, the equivalence classes are conserved. The designation of the classes may have changed, but this is irrelevant since we only need to identify the most likely.
    \item To conserve the probability of all error chains, the error model has to be adjusted such that the probability of the four possible link errors correspond to those for the triggered link stabilizer if there were such, following \tabref{table:prior-update-rule}.
    \item Finally, use the maximum-likelihood decoder to decode the corresponding YZZY code, with appropriately locally modified error rates. With this construction, identifying the most likely equivalence class for the syndrome on the latter also identifies the most likely class for original syndrome of the \xyz code.
    \item The construction of an error-recovery operator on the \xyz code based on one for the YZZY code simply follows from the mapping in \tabref{table:prior-update-rule}, differentiated by the existence of triggered links or not in the original syndrome.
\end{enumerate}

\end{document}